# Pressure dependent topological, superconducting, optoelectronic and thermophysical properties of Ta$_2$Se chalcogenide: Theoretical insights

Tauhidur Rahman, Jubair Hossan Abir, Sourav Kumar Sutradhar,
Sraboni Saha Moly, Mst. Maskura Khatun, Md. Asif Afzal and Saleh Hasan Naqib*

Department of Physics, University of Rajshahi, Rajshahi 6205, Bangladesh
*Corresponding author; Email: salehnaqib@yahoo.com

**Abstract**

Tetragonal Ta$_2$Se is a layered, Ta-rich (metal-rich) chalcogenide that departs from conventional MX$_2$ transition-metal dichalcogenides by hosting dense Ta-Ta networks capped by Se square-net layers. Here, we present a unified first-principles investigation of hydrostatic-pressure tuning in Ta$_2$Se from 0 to 10 GPa, connecting the structural response, mechanical stability, thermophysical indicators, bonding evolution, electronic and optical behavior, lattice dynamics and superconductivity within a single framework. Pressure induced a smooth, monotonic lattice contraction with an overall volume reduction of about 9.9% at 10 GPa and a stronger compressibility along the *c*-axis, consistent with pressure-enhanced interlayer coupling in the layered stacking. All independent elastic constants satisfied the Born-Huang stability conditions throughout the studied range and increased systematically with pressure, leading to enhanced bulk, shear and Young's moduli and a modest rise in the predicted hardness. Pugh's ratio and Poisson's ratio consistently indicate ductile, predominantly metallic bonding, whereas elastic anisotropy remains weak to moderate without signatures of instability. The derived thermophysical descriptors corroborate a pressure-stiffened lattice: density increases, Debye temperature rises, melting temperature is elevated and minimum thermal conductivity increases, whereas the Grüneisen parameter remains within a narrow window, suggesting no anomalous anharmonic softening. Bond population metrics and electron-density-difference analysis revealed a mixed metallic-covalent bonding picture dominated by a robust Ta-Ta metallic backbone, accompanied by pressure-strengthened Ta-Se hybridization. Electronic-structure calculations show persistent metallicity under compression; pressure broadens bands, reduces N($E_F$), reshapes the Fermi surface and points to a possible Lifshitz-type reconstruction without symmetry breaking. The optical response remained metallic with Drude-like low-energy behavior and pressure-tunable spectral features. The phonon dispersions exhibit no imaginary modes (checked at representative pressures), confirming dynamical stability. Electron-phonon coupling calculations classify Ta$_2$Se as a weak-coupling, phonon-mediated superconductor with $T_c \approx 3.9$ K, consistent with available experiments and establish pressure as a practical control knob for stability and superconductivity-relevant descriptors in this metal-rich layered platform.

# 1   Introduction

Chalcogenides are compounds that include at least one group-16 element (S, Se, or Te) bonded to a more electropositive partner, typically a metal. Chalcogenides form one of the most versatile material families in condensed-matter physics and functional materials because small changes in bonding, coordination and dimensionality can move the ground state between semiconducting, topological and superconducting regimes. In particular, layered chalcogenides offer a tunable balance between strong in-plane bonding and weak interlayer (van der Waals) coupling, making them ideal candidates for controlling properties by thickness, strain, chemical substitution, and pressure. Within this family, binary chalcogenides remain central because they provide clean structure-property testbeds while already enabling key technologies. They are widely used as 'model' functional materials because a simple chemistry can deliver strongly different ground states (semiconducting, metallic, topological) and device-relevant responses. Metal-rich chalcogenides are compositions with extra metal compared with simple 1:1 or 1:2 ratios. Metal-rich chalcogenides, often called subchalcogenides, are compounds in which metal atoms outnumber the chalcogen atoms (S, Se, or Te) and their bonding reflects an interplay of metal-metal and metal-chalcogen interactions [1-3]. Because the composition is metal-heavy, the electronic states near the Fermi level are typically dominated by metal $d$-orbitals arising from dense metal-metal networks. Structurally, many of these materials can be viewed as the anti-TMD counterparts of dichalcogenides: several-atom-thick metal slabs capped by chalcogen layers instead of single metal layers sandwiched between chalcogens. In many of these compounds, the metal atoms form slabs or clusters that are terminated by chalcogen layers, an anti-format of the better-known $MX_2$ dichalcogenides. Thus, the electronic states near the Fermi level are dominated by metal $d$-orbitals rather than chalcogen $p$-states. Metal-rich chalcogenides where excess metal introduces strong metal-metal bonding offer high conductivity, tunable electronic structures and robust layered frameworks [4-6]. These traits enable applications from low-temperature superconducting circuitry and high-field conductors to energy technologies such as multivalent-ion battery cathodes and mid to high-temperature thermoelectrics. Their active surfaces and mixed bonding make them promising electrocatalysts (e.g., for hydrogen evolution) and conductive supports, while the strong spin-orbit coupling in Se/Te-rich members opens avenues for spintronics and topological-device concepts. Due to their sharp response to composition, intercalation and pressure, they provide practical levers for device optimization [7-10].

The most widely known layered chalcogenides are chalcogen-rich (e.g., $MX_2$). $Ta_2Se$, in contrast, is metal-rich (Ta-rich), placing it in a less explored regime where metal-metal bonding networks and unusual layer motifs can dominate the physics and is structurally distinct from standard $MX_2$ TMDs. It crystallizes in a layered tetragonal structure with space group P4/*nmm* (No. 129) [11], comprising four close-packed Ta layers sandwiched between two square-net Se layers, forming repeated Se-Ta-Ta-Ta-Ta-Se building blocks. Experimentally, the modern baseline for $Ta_2Se$ is the establishment of bulk superconductivity in this material. $Ta_2Se$ is reported as a bulk superconductor with $T_c \approx 3.8$ K by Gui et al. [12]. Thermodynamic analysis indicates weak-coupling BCS behavior with an electron-phonon coupling constant around $\lambda \approx 0.61$. Electronic-structure calculations show that Ta $d$-states dominate at the Fermi level and reveal flat bands and van Hove singularity near $E_F$ that are sensitive to spin-orbit coupling, a feature that can support superconductivity. Low-temperature diffraction found no evidence of a charge-density wave above 100 K in $Ta_2Se$, underlining its contrast with canonical TMD behavior such as in $TaSe_2$. Recently, Lee et al. [13] identified phonon vibrational modes of layered $Ta_2Se$ by Raman spectroscopy supported by DFPT calculations, providing essential lattice-dynamical benchmarks for a superconducting, layered, metal-rich chalcogenide.

The second pillar of metal-rich chalcogenide superconductivity is the Chevrel-phase family $M_xMo_6X_8$ (X = S/Se/Te), long known for $T_c$ ~ 12-15 K and unusual normal-state and vortex physics [14]. Flagship members $PbMo_6S_8$ and $SnMo_6S_8$ exhibit very high upper critical fields ($H_{c2} \approx$ 40-60 T) coupled with $T_c$ values in the mid-teens. First-principles SCDFT work on $PbMo_6S_8$ confirms phonon-mediated pairing and shows how proximity to structural instabilities and Coulomb effects shape the superconducting state [15]. A practical way to organize superconducting metal-rich chalcogenides is by their structural motifs, because structure strongly governs the bands near $E_F$ and the resulting pairing tendencies [12, 14].

Pressure is a particularly powerful tuning parameter for layered chalcogenides because it modifies interatomic distances without introducing chemical disorder, directly changing orbital overlap and phonon spectra [16]. In closely related Ta-based layered systems, pressure can strongly reshape the competition between charge-density-wave order and superconductivity [17-19]. $Ta_2Se$ lacks a systematic pressure-dependent mapping of structure-electronics-lattice dynamics that can connect compressibility and anisotropy to changes in density of states near $E_F$, phonon stiffness and superconductivity-relevant descriptors. This gap is especially relevant because $Ta_2Se$ exhibits Ta $d$-dominated states at $E_F$ and van Hove related DOS features that are strengthened by SOC, suggesting that pressure could modify the near-$E_F$ band topology and thereby alter superconductivity descriptors, however such effects remain largely unquantified for $Ta_2Se$ [12].

This work aims to develop a coherent, pressure-dependent understanding of tetragonal $Ta_2Se$ from 0-10 GPa by connecting its structural evolution, mechanical stability, thermophysical indicators, bonding and charge redistribution, electronic and optical responses, lattice dynamics and superconductivity within a single first-principles framework. In particular, we determine how hydrostatic pressure modifies (i) the equilibrium geometry and compressibility, (ii) elastic constants and derived mechanical descriptors, (iii) selected thermophysical quantities derived from elastic data, (iv) chemical bonding and charge rearrangement via bond population and electron-density-difference analyses, (v) band structure and density of states, (vi) optical spectra from the dielectric response, (vii) dynamical stability from phonon dispersions and (viii) the superconducting critical temperature by computing $T_c$ from electron-phonon coupling parameters.

## 2    Computational Methodology

First-principles calculations were performed within the framework of density functional theory (DFT) [20, 21] as contained within the Cambridge Serial Total Energy Package (CASTEP) code [22, 23] with the Vanderbilt-type ultrasoft pseudopotential [24] and the VASP package [25, 26] with the projector augmented wave (PAW) method [27, 28]. The exchange-correlation energy was treated using the generalized gradient approximation (GGA), specifically the Perdew-Burke-Ernzerhof for solids (PBEsol) [29] scheme. The interactions of valence electrons and the ion cores were modeled with the Vanderbilt-type ultrasoft pseudopotential [24]. The valence electron configurations for Ta and Se elements are taken as $5s^2\ 5p^6\ 5d^3\ 6s^2$ and $4s^2\ 4p^4$, respectively.

Geometry optimization of $Ta_2Se$ was performed using the BFGS minimization scheme [30] to reduce the total energy and residual forces. A plane-wave cutoff energy of 550 eV was used for all pressures, ensuring good convergence. Brillouin-zone sampling employed a Monkhorst-Pack $k$-point grid of $12 \times 12 \times 4$ [31], which adequately satisfied the total-energy convergence criteria. The calculation tolerances were set to energy < $10^{-5}$ eV/atom, maximum displacement

< $10^{-3}$ Å, maximum ionic force < 0.03 eV Å$^{-1}$, maximum stress < 0.05 GPa, and a smearing width of 0.1 eV, including finite basis set corrections [32]. For the Fermi-surface construction, a denser *k*-point mesh of $32 \times 32 \times 11$ was applied.

The elastic constants of Ta$_2$Se were calculated using the stress-strain method [33]. Due to the crystal symmetry, a tetragonal crystal is characterized by six independent second-order elastic coefficients, which are $C_{11}$, $C_{33}$, $C_{44}$, $C_{66}$, $C_{12}$ and $C_{13}$. Using the independent elastic constants $C_{ij}$, we computed the macroscopic (polycrystalline) elastic moduli, bulk modulus (B), shear modulus (G), and Young's modulus (Y) via the Voigt-Reuss-Hill (VRH) averaging scheme, where applicable [34, 35].

The optical response of Ta$_2$Se as a function of frequency is obtained from its complex dielectric function, $\varepsilon(\omega) = \varepsilon_1(\omega) + i\varepsilon_2(\omega)$. The imaginary part, $\varepsilon_2(\omega)$, is specifically calculated using the following relation:

$$\varepsilon_2(\omega) = \frac{2e^2\pi}{\Omega\varepsilon_0} \sum_{k,v,c} |\langle \psi_k^c | \hat{u} \cdot \vec{r} | \psi_k^v \rangle|^2 \delta(E_k^c - E_k^v - E) \qquad (1)$$

Here, $\Omega$ is the unit-cell volume, $\omega$ is the photon frequency, $\varepsilon_0$ is the vacuum permittivity, e is the electron charge, r is the position vector, u sets the polarization of the incident field and $\psi_k^c$ and $\psi_k^v$ are the conduction and valence band wavefunctions at wave vector k. The expression is evaluated directly from the calculated electronic band structure, and the real part of the dielectric function, $\varepsilon_1(\omega)$, is obtained from $\varepsilon_2(\omega)$ using the Kramers-Kronig relation. Once the dielectric function is determined, the remaining optical constants-refractive index n($\omega$), absorption coefficient α($\omega$), energy loss function L($\omega$), reflectivity R($\omega$), and optical conductivity σ($\omega$) can be obtained directly using standard formalism [36, 37].

To examine the bonding characteristics of Ta$_2$Se, the plane-wave states were projected onto a linear combination of atomic orbitals (LCAO) basis sets [38, 39], enabling a Mulliken bond population analysis [40]. In this approach, the Mulliken density operator is expressed in an atomic (or quasi-atomic) basis and then used to evaluate bond population analysis:

$$P_{\mu\nu}^M(g) = \sum_{g'}\sum_{\nu'} P_{\mu\nu'}(g') S_{\nu'\nu}(g-g') = L^{-1} \sum_k e^{-ikg} (P_k S_k)_{\mu\nu'} \qquad (2)$$

and the net charge on atom A can be defined as,

$$Q_A = Z_A - \sum_{\mu \in A} P_{\mu\mu}^M(0) \qquad (3)$$

Where, $Z_A$ is the nuclear charge.

To examine how pressure and temperature affect the thermodynamic, elastic, and structural behaviors, we used the energy-volume (E-V) data fitted with the third-order Birch-Murnaghan equation of state [41]. The CASTEP code was used to compute elastic, optical, and bonding properties; VASP was used for detailed electronic band-structure analysis.

To investigate the superconducting properties, density functional theory calculations were carried out using the Quantum ESPRESSO (QE) [42] package. The PAW-type pseudopotentials, constructed within the PBE [27] exchange–correlation framework, were taken from the PSLIBRARY [43]. The plane-wave kinetic energy cutoffs were set at 80 Ry for the wavefunctions and 640 Ry for the charge density. Phonon dispersion relations were obtained using density functional perturbation theory (DFPT). For the evaluation of electron–phonon coupling, a $19 \times 19 \times 7$ Monkhorst-Pack *k*-point mesh along with a $3 \times 3 \times 3$ *q*-point grid was employed to ensure sufficient accuracy.

## 3 Results and Discussion

### 3.1 Structural Properties

The crystal structure of Ta$_2$Se is Tetragonal with space group P4/*nmm* (No. 129), which is a non-symmorphic with high symmetry layered material. **Figure 1** shows the crystal structure of Ta$_2$Se. The Ta and Se atoms occupy the following Wyckoff position in unit cell: Ta1 at (0.25 0.25 0.0776), Ta2 at (0.25 0.25 0.7553) and Se at (0.25 0.25 0.3665) [11]. The unit cell contains four close-packed Ta layers between two square Se nets, forming Se-Ta-Ta-Ta-Ta-Se slabs with two formula units.

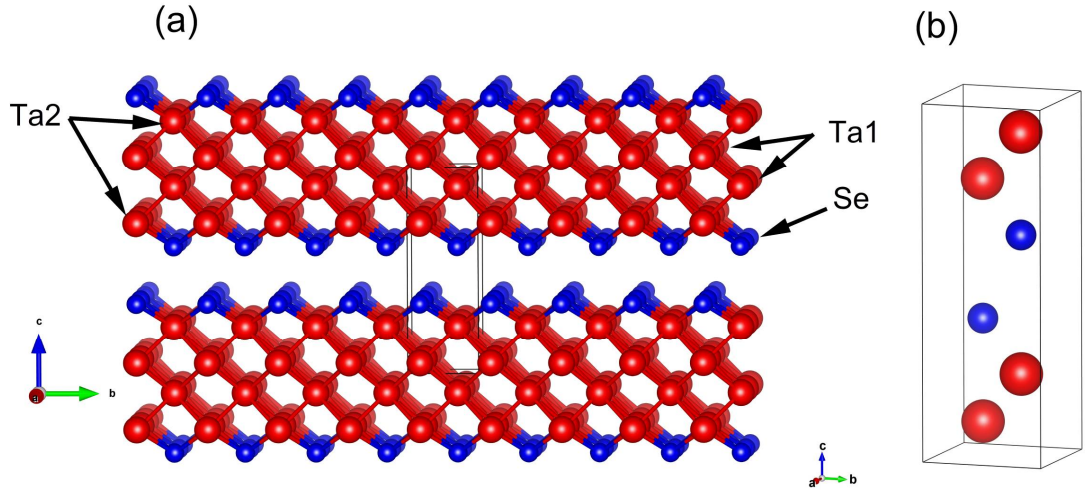

**Figure 1.** (a) Layered and (b) three-dimensional (3D) schematic of the crystal structure of the tetragonal Ta$_2$Se compound.

The lattice constants and unit-cell volume of Ta$_2$Se decrease monotonically with pressure. The optimized volume drops from 111.787 Å³ at 0 GPa to 102.285 Å³ at 10 GPa ($\approx$ 9.86%). Normalized $a/a_0$ and $c/c_0$ (where $a_0$, $c_0$, and $V_0$ represent the structural parameters and unit cell volume at 0 GPa) both contract with pressure. However, the compression is more pronounced along $c$ and a slope change occurs around 4 GPa.

**Table 1.** Pressure-dependent optimized lattice parameters ($a = b$, and $c$ in Å), cell volume $V$ in Å³ and cohesive energy ($E_{coh}$ in eV/atoms) of Ta$_2$Se.

| Pressure (GPa) | $a = b$ | $c$ | $c/a$ | $a/a_0$ | $c/c_0$ | $V$ | $V/V_0$ | $E_{coh}$ | Remarks |
|---|---|---|---|---|---|---|---|---|---|
| - | 3.375 | 9.832 | 2.913 | | | 111.993 | | - | Expt. [11] |
| 0 | 3.365 | 9.874 | 2.934 | 1 | 1 | 111.787 | 1 | -10.31 | This work |
| 2 | 3.355 | 9.637 | 2.872 | 0.997 | 0.976 | 108.545 | 0.971 | -10.01 | |
| 4 | 3.345 | 9.469 | 2.831 | 0.994 | 0.959 | 105.974 | 0.948 | -9.68 | |
| 6 | 3.335 | 9.410 | 2.823 | 0.991 | 0.953 | 104.633 | 0.936 | -9.46 | |
| 8 | 3.325 | 9.360 | 2.815 | 0.988 | 0.948 | 103.403 | 0.925 | -9.25 | |
| 10 | 3.314 | 9.311 | 2.809 | 0.985 | 0.943 | 102.285 | 0.915 | -9.04 | |

Thermodynamic stability was assessed from the cohesive energy per atom, following refs. [44, 45]. The cohesive energy was evaluated as:

$$E_{coh} = \frac{E_{Ta_2Se} - 4E_{Ta} - 2E_{Se}}{6} \quad (4)$$

Here, $E_{Ta_2Se}$ is the total energy per unit cell of Ta$_2$Se and $E_{Ta}$ and $E_{Se}$ are the total energies of Ta and Se in their solid states. Ta$_2$Se shows negative cohesive (binding) energy at all pressures, confirming thermodynamic stability [46]. Its pressure dependence is smooth, with no abrupt anomaly over the investigated pressure regime.

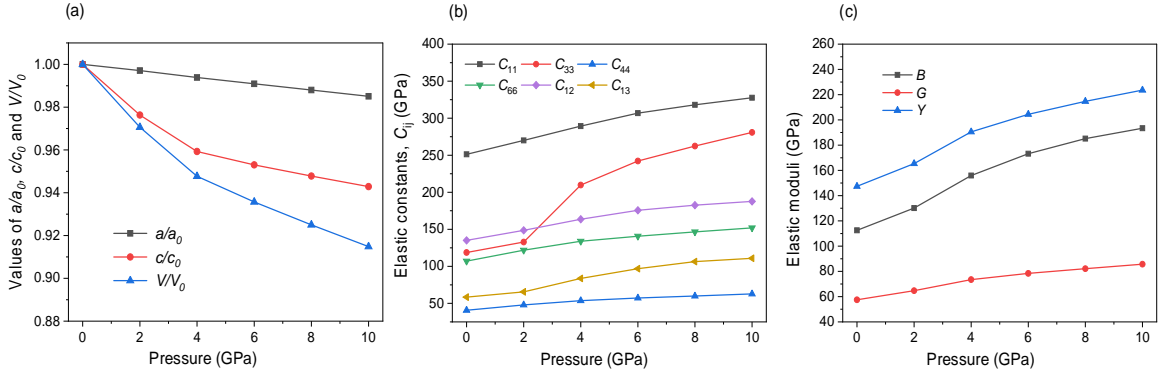

**Figure 2.** (a) Normalized parameters $a/a_0$, $b/b_0$, $c/c_0$, $V/V_0$, (b) Single crystal elastic constants and (c) Elastic moduli of Ta$_2$Se under different pressures.

## 3.2 Elastic and Mechanical Properties

Elastic constants quantify stiffness and mechanical stability under applied stress. A tetragonal crystal has six independent elastic constants ($C_{11}$, $C_{33}$, $C_{44}$, $C_{66}$, $C_{12}$ and $C_{13}$); the calculated values for Ta$_2$Se are listed in **Table 2**. All constants are positive and satisfy the Born–Huang stability criteria [47]: $C_{11} > 0$, $C_{33} > 0$, $C_{44} > 0$, $C_{66} > 0$, $(C_{11} - C_{12}) > 0$, $(C_{11} + C_{33} - 2C_{13}) > 0$, $\{2(C_{11} + C_{12}) + C_{33} + 4C_{13}\} > 0$. The positive values of all the elastic constants for Ta$_2$Se confirm its stability against static deformation.

**Table 2.** Computed elastic constants, $C_{ij}$ (GPa), tetragonal shear modulus, $C'$ (GPa), Cauchy pressure, $C'' = (C_{12} - C_{44})$ (GPa) and Kleinman parameter ($\zeta$) for Ta$_2$Se.

| Pressure (GPa) | $C_{11}$ | $C_{33}$ | $C_{44}$ | $C_{66}$ | $C_{12}$ | $C_{13}$ | $C'$ | $C''$ | $\zeta$ | Remarks |
|---|---|---|---|---|---|---|---|---|---|---|
| 0 | 251.21 | 118.68 | 40.72 | 107.07 | 135.02 | 58.38 | 58.095 | 94.30 | 0.656 | |
| 2 | 270.04 | 132.85 | 47.95 | 121.81 | 148.66 | 65.50 | 60.690 | 100.71 | 0.667 | |
| 4 | 289.31 | 209.79 | 53.70 | 133.91 | 163.61 | 83.70 | 62.850 | 109.91 | 0.679 | This work |
| 6 | 306.88 | 242.18 | 57.23 | 140.66 | 175.72 | 96.87 | 65.580 | 118.49 | 0.685 | |
| 8 | 318.08 | 262.40 | 59.86 | 146.46 | 182.55 | 106.34 | 67.765 | 122.69 | 0.686 | |
| 10 | 327.66 | 280.80 | 62.73 | 151.92 | 187.68 | 110.83 | 69.990 | 124.95 | 0.685 | |

$C_{11}$ and $C_{33}$ describe resistance to uniaxial compression along the $a$-and $c$-axes, respectively, and both increase with pressure. The smaller initial $C_{33}$ indicates a more compressible c-axis at ambient conditions, while its rapid rise reflects strong stiffening under compression. Shear constants $C_{44}$ and $C_{66}$ also exhibit an upward trend and $C_{66} > C_{44}$ implies greater shear rigidity within the basal plane. The coupling terms $C_{12}$ and $C_{13}$ grow with pressure, consistent with reduced structural compliance and stronger interlayer coupling.

The tetragonal shear constant, $C' = \frac{C_{11} - C_{12}}{2}$ increases from 58.095 GPa at 0 GPa to 69.99 GPa at 10 GPa, indicating enhanced resistance to shear distortion and improved mechanical stability at higher pressure.

The Cauchy pressure, defined as $C'' = C_{12} - C_{44}$, remains positive and rises from 94.30 GPa to 124.95 GPa with pressure. According to Pettifor's rule [48], this supports predominantly metallic bonding and exhibits high level of ductility, suggesting that Ta$_2$Se retains and strengthens its metallic bonding nature and ductility under compression.

The Kleinman internal strain parameter ζ [49] increases slightly from 0.656 to 0.685 with pressure, indicating a moderate shift toward bond-stretching mechanisms relative to bond-bending mechanisms during internal relaxation under strain [46].

The standard formulae are used to estimate the Hill values of bulk modulus (BH), shear modulus (GH) (obtained using the Voigt-Reuss-Hill (VRH) method), and Young's modulus (Y), which are detailed in Refs. [50-52]. **Table 3** summarizes the pressure evolution the bulk modulus (B), shear modulus (G), Young's modulus (Y), Pugh's ratio (G/B), Poisson's ratio (ν), Machinability index $\mu_M$ and Vickers hardness $H_V$ from 0 to 10 GPa. The bulk modulus (B) indicates resistance to volume change, while the shear modulus (G) reflects resistance to shear. For Ta$_2$Se, the lower G value compared to B suggests that its mechanical strength is limited by shear deformation. A high shear modulus typically indicates strong directional atomic bonding between atoms [53, 54]. The stiffness (or resistance) of an elastic material to a change in length is measured by its Young's modulus [55]. When a material's Young's modulus is high, it is considered stiff. The relatively high Young's modulus (Y) indicates that Ta$_2$Se is expected to remain reasonably stiff across the investigated pressure range. The Young's modulus transitions from moderate to high with applied pressure.

**Table 3.** The calculated isotropic bulk modulus B (GPa), shear modulus G (GPa), Young's modulus Y (GPa), Pugh's indicator G/B, Poisson's ratio ν, Machinability index $\mu_M$ and Vickers hardness $H_V$ (GPa) for the Ta$_2$Se compound subjected to varying pressures.

| Pressure (GPa) | $B_V$ | $B_R$ | $B_H$ | $G_V$ | $G_R$ | $G_H$ | Y | G/B | ν | $\mu_M$ | $H_V$ | Remarks |
|---|---|---|---|---|---|---|---|---|---|---|---|---|
| 0 | 124.96 | 100.03 | 112.50 | 62.325 | 52.70 | 57.51 | 147.41 | 0.511 | 0.282 | 2.763 | 8.373 | |
| 2 | 136.91 | 123.37 | 130.14 | 69.76 | 59.62 | 64.69 | 165.35 | 0.497 | 0.287 | 2.714 | 9.132 | |
| 4 | 161.15 | 150.64 | 155.90 | 78.50 | 68.50 | 73.50 | 190.55 | 0.471 | 0.296 | 2.903 | 9.982 | This work |
| 6 | 177.20 | 169.30 | 173.25 | 83.45 | 73.37 | 78.41 | 204.4 | 0.453 | 0.303 | 3.027 | 10.279 | |
| 8 | 188.00 | 182.40 | 185.20 | 87.40 | 76.91 | 82.15 | 214.70 | 0.443 | 0.307 | 3.094 | 10.582 | |
| 10 | 193.20 | 187.60 | 190.40 | 91.10 | 80.30 | 85.70 | 223.60 | 0.450 | 0.304 | 3.035 | 11.183 | |

Pugh's ratio *G/B*, represents a solid's brittleness or ductility [56, 57]. A material is ductile if the value of *G/B* is less than 0.57; otherwise, it would be brittle. Ta$_2$Se satisfies this condition at all investigated pressures [**Table 3; Figure 3**(a)], indicating ductility throughout 0 to 10 GPa.

Poisson's ratio (ν) generally ranges from -1 to 0.50. The critical Poisson's ratio is 0.26. Materials with values above this threshold tend to show ductile behavior, while those below it is generally brittle. Poisson's ratio is also closely linked to the nature of interatomic bonding in solids [58]. Also, ν = 0.25-0.50 usually means central-force bonding dominates [59]. Moreover, Poisson's ratio also provides insight into the relative contributions of covalent and ionic bonding within a compound [60]. As shown in **Table 3** and **Figure 3**(b), $Ta_2Se$ stays in 0.25-0.50 and remains > 0.26 under pressures, so it is ductile and mainly governed by central-force bonding. Thus, the same conclusion can be drawn from both ν and G/B.

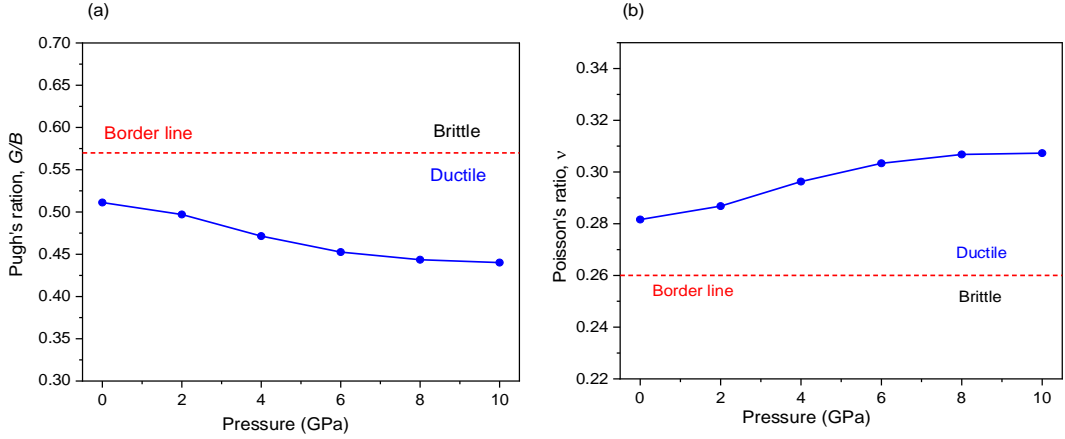

**Figure 3.** (a) Pugh's ratio and (b) Poisson's ratio of $Ta_2Se$ under pressure.

Machinability describes with which a material can be cut or shaped during machining processes. The machinability index is defined as: $\mu_M = B/C_{44}$ [61]. It is also a useful parameter for assessing the plasticity and dry lubrication behavior of a solid [62-64]. A higher value of $\mu_M$ indicates that the material is easier to machine. The increase in $\mu_M$ with pressure suggests that $Ta_2Se$ becomes more machinable as pressure is applied.

Vickers Hardness measures a material's resistance to indentation. It rises from 8.373 GPa at 0 GPa to 11.183 GPa at 10 GPa, indicating moderate hardness that is further enhanced by pressure [65]. The increase reflects pressure-enhanced bonding and greater resistance to indentation.

### 3.3 Elastic Anisotropy

Elastic anisotropy describes how a crystal's mechanical response varies depending on the crystallographic direction [66]. Such anisotropy affects crack propagation, dislocation motion, plastic deformation, and phonon behavior, among other processes [54, 67]. A clear explanation of anisotropy factors makes it easier to understand how solids respond to stress under various conditions [68]. The shear anisotropy factors can be determined using the following equations: [69]. The shear anisotropy factor for the {100} planes, measured between the ⟨011⟩ and ⟨010⟩ directions,

$$A_1 = \frac{4C_{44}}{C_{11} + C_{33} - 2C_{13}} \quad (5)$$

For the {010} shear planes, measured between the ⟨101⟩ and ⟨001⟩ directions:

$$A_2 = \frac{4C_{55}}{C_{22} + C_{33} - 2C_{23}} \tag{6}$$

For the {001} shear planes, measured between the ⟨101⟩ and ⟨001⟩ directions:

$$A_3 = \frac{4C_{66}}{C_{11} + C_{22} - 2C_{12}} \tag{7}$$

The shear anisotropy factor $A_1$, $A_2$ and $A_3$ should be unity for isotropic system. The calculated $A_1$, $A_2$ and $A_3$ values indicate moderate anisotropy in Ta$_2$Se. The largest deviation occurs in $A_3$, reaching 2.171 at 10 GPa, while $A_1$ equals $A_2$ at all investigated pressures.

The universal anisotropy index $A^U$, the equivalent Zener anisotropy measure $A^{eq}$, anisotropy in compressibility, $A^B$ and anisotropy in shear $A^G$ (or $A^C$) were evaluated using standard formulas valid for crystals of any symmetry: [67, 70, 71].

$$A^U = \frac{5G_V}{G_R} + \frac{B_V}{B_R} - 6 \geq 0 \tag{8}$$

$$A^{eq} = \left(1 + \frac{5}{12}A^U\right) + \left(\left(1 + \frac{5}{12}A^U\right)^2 - 1\right)^{0.5} \tag{9}$$

$$A^B = \frac{B_V - B_R}{B_V + B_R} \tag{10}$$

$$A^G \text{ (or } A^C\text{)} = \frac{G_V - G_R}{2G^H} \tag{11}$$

The universal anisotropy factor $A^U$, can also be explained as a generalization of the Zener anisotropy factor and defined by the above equation. Isotropic crystal, when $A^U = 0$; Anisotropic if $A^U > 0$. The Ta$_2$Se compound exhibits elastic anisotropy, as indicated by the calculated values of $A^U$. The material shows $A^U = 1.162$ at 0 GPa and a decreasing trend with increasing pressure but the decline rate significantly slows as pressure rises.

The Zener anisotropy factor, $A^{eq} = 1$ denotes an isotropic state. Ta$_2$Se has $A^{eq} = 2.581$ at 0 GPa and 2.112 at 10 GPa, showing a slight decrease in anisotropy with pressure.

$A^B = 0$ and $A^G$ (or $A^C$) $= 0$ corresponds to the elastic isotropy, while any deviation greater than zero quantify anisotropy. The data in **Table 4** reveal a unique behavior at ambient pressure, where the bulk anisotropy ($A^B$) is higher than the shear anisotropy ($A^G$), however, shear anisotropy becomes dominant at higher pressures [**Figure 4**(b)]. Notably, $A^B$ consistently indicates the lowest overall anisotropy for Ta$_2$Se compared to the other anisotropy measures when pressure is increased.

The universal log-Euclidean index $A^L$ can be calculated using equation (12), where the Reuss and Voigt values of $C_{44}$ are provided below [70, 72].

$$A^L = \sqrt{\left[\ln\left(\frac{B^V}{B^R}\right)\right]^2 + 5\left[\ln\left(\frac{C_{44}^V}{C_{44}^R}\right)\right]^2} \tag{12}$$

$$C_{44}^R = \frac{5}{3} \frac{C_{44}(C_{11}-C_{12})}{3(C_{11}-C_{12})+4C_{44}} \tag{13}$$

and

$$C_{44}^V = C_{44}^R + \frac{3}{5} \frac{(C_{11}-C_{12}-2C_{44})^2}{3(C_{11}-C_{12})+4C_{44}} \tag{14}$$

The expression for $A^L$ applies to all crystal symmetries and is closely related to the universal anisotropy index $A^U$. For a perfectly isotropic crystal, $A^L$ equals 0. For Ta$_2$Se, the calculated $A^L$ values indicate a generally low anisotropy across all investigated pressures. Because large $A^L$ is often associated with layered structures [46, 70], the relatively small $A^L$ suggests Ta$_2$Se is not strongly lamellar.

**Table 4.** Shear anisotropic factors ($A_1$, $A_2$ and $A_3$), universal log-Euclidean index ($A^L$), the universal anisotropy index ($A^U$), equivalent Zener anisotropy measure ($A^{eq}$), anisotropy in shear ($A_G$ or $A_C$) and anisotropy in compressibility ($A_B$), linear compressibilities ($\beta_a$ and $\beta_c$) and their ratio ($\beta_c/\beta_a$) for Ta$_2$Se.

| Pressure (GPa) | $A_1$ | $A_2$ | $A_3$ | $A^U$ | $A^{eq}$ | $A^B$ | $A^G$ | $A^L$ | $\beta_a$ | $\beta_c$ | $\frac{\beta_c}{\beta_a}$ | Remarks |
|---|---|---|---|---|---|---|---|---|---|---|---|---|
| 0 | 0.643 | 0.643 | 1.843 | 1.162 | 2.581 | 0.111 | 0.084 | 0.297 | 0.0015 | 0.0069 | 4.469 | |
| 2 | 0.705 | 0.705 | 2.007 | 0.960 | 2.380 | 0.052 | 0.078 | 0.136 | 0.0014 | 0.0061 | 4.272 | |
| 4 | 0.647 | 0.647 | 2.131 | 0.799 | 2.215 | 0.034 | 0.068 | 0.078 | 0.0015 | 0.0035 | 2.264 | This work |
| 6 | 0.644 | 0.644 | 2.145 | 0.733 | 2.145 | 0.023 | 0.064 | 0.054 | 0.0015 | 0.0029 | 1.988 | |
| 8 | 0.651 | 0.651 | 2.161 | 0.713 | 2.123 | 0.015 | 0.064 | 0.039 | 0.0014 | 0.0026 | 1.845 | |
| 10 | 0.649 | 0.649 | 2.171 | 0.702 | 2.112 | 0.015 | 0.063 | 0.035 | 0.0014 | 0.0024 | 1.728 | |

The linear compressibility of a tetragonal compound along a and c axis ($\beta_a$ and $\beta_c$) are calculated from [73]:

$$\beta a = \frac{C_{33} - C_{13}}{D} \quad and \quad \beta c = \frac{C_{11} + C_{12} - 2C_{13}}{D} \tag{15}$$

$$with, \ D = (C_{11} + C_{12}) C_{33} - 2(C_{13})^2$$

The calculated values in **Table 4** show anisotropic compressibility along a and c, consistent with directional elastic response. However, compared with other measures, linear compressibility indicates negligible anisotropy in Ta$_2$Se.

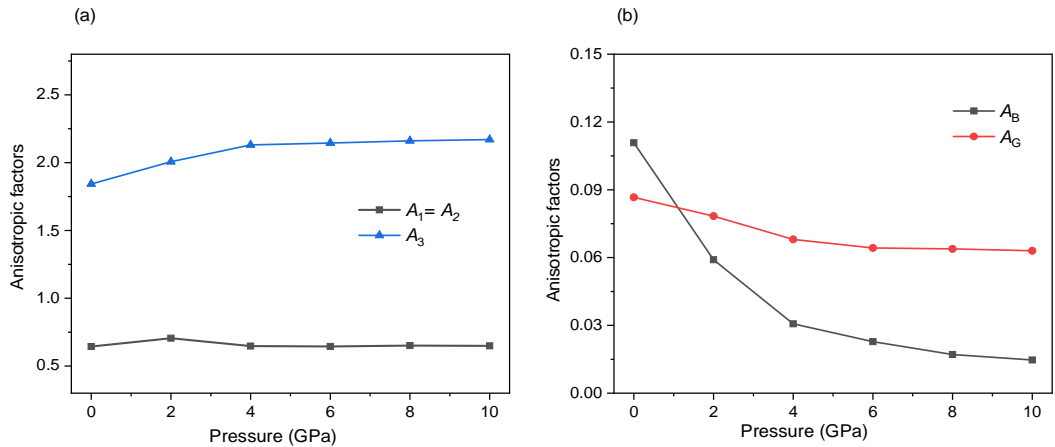

**Figure 4.** Elastic anisotropy indices and their pressure dependence for the Ta$_2$Se compound.

For an isotropic solid, 3D surfaces of Young's modulus ($Y$), shear modulus ($G$), linear compressibility ($\beta$), and Poisson's ratio ($v$) are spherical; deviations from a sphere indicate anisotropy. Directional variations of these properties for Ta$_2$Se were visualized using ELATE tool [74, 75] [**Figure 5**-**8**]. The matrix is imported into ELATE to visualize the degree of anisotropy of the Young's modulus ($Y$), shear modulus ($G$), Poisson's ratio ($v$), and compressibility ($\beta$) of the Ta$_2$Se material. Green and blue regions indicate the minimum and maximum values, while red regions show negative values. Only small deviations from sphericity are observed across all pressures, which is consistent with weak-to-moderate anisotropy. **Table 5** displays the upper and lower bounds for these variables from 0 to 10 GPa.

**Table 5.** The minimum and maximum limits of Young's modulus, $Y$ (GPa), linear compressibility, $\beta$, Shear modulus, $G$ (GPa) and Poisson's ratio, $v$ of the Ta$_2$Se compound

| Pressure (GPa) | Y | | $A_Y$ | β | | $A_\beta$ | G | | $A_G$ | v | | $A_v$ | Remarks |
|---|---|---|---|---|---|---|---|---|---|---|---|---|---|
| | $Y_{min}$ | $Y_{max}$ | | $\beta_{min}$ | $\beta_{max}$ | | $G_{min}$ | $G_{max}$ | | $v_{min}$ | $v_{max}$ | | |
| 0 | 99.68 | 259.37 | 2.60 | 1.54 | 6.90 | 4.47 | 40.72 | 107.07 | 2.63 | 0.131 | 0.478 | 3.66 | |
| 2 | 112.30 | 288.65 | 2.57 | 1.43 | 6.11 | 4.27 | 47.95 | 121.81 | 2.54 | 0.133 | 0.489 | 3.66 | |
| 4 | 144.63 | 316.26 | 2.19 | 1.55 | 3.52 | 2.26 | 53.07 | 133.90 | 2.52 | 0.125 | 0.489 | 3.92 | This work |
| 6 | 157.06 | 332.02 | 2.11 | 1.48 | 2.94 | 1.99 | 57.23 | 140.64 | 2.46 | 0.129 | 0.502 | 3.89 | |
| 8 | 165.29 | 344.54 | 2.08 | 1.45 | 2.57 | 1.76 | 59.85 | 146.46 | 2.45 | 0.128 | 0.514 | 4.01 | |
| 10 | 172.52 | 357.45 | 2.07 | 1.43 | 2.47 | 1.73 | 62.73 | 151.92 | 2.42 | 0.127 | 0.510 | 4.01 | |

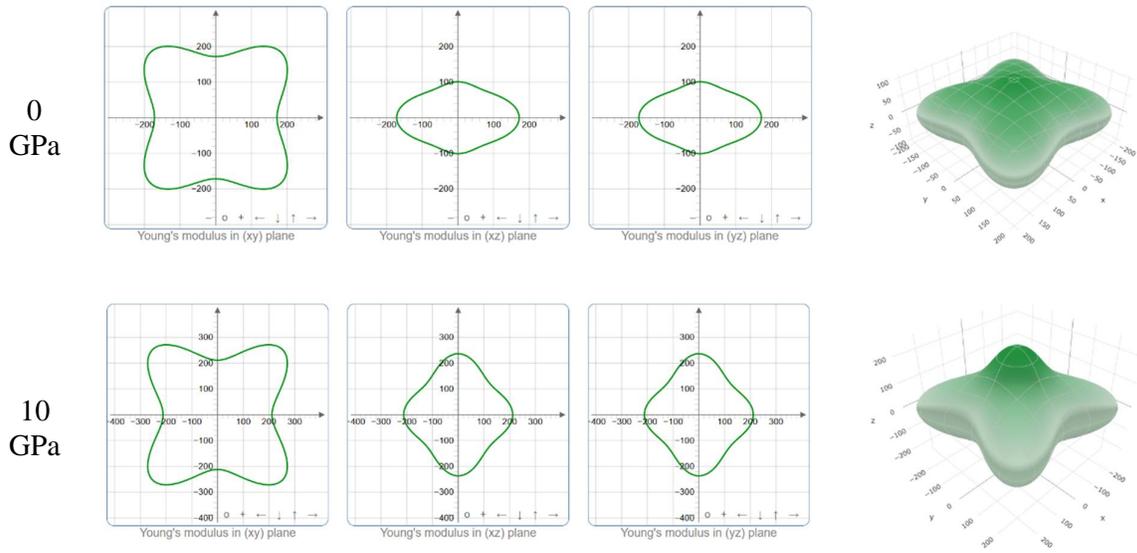

**Figure 5.** Directional variation in Young's modulus (*Y*) of Ta$_2$Se compound.

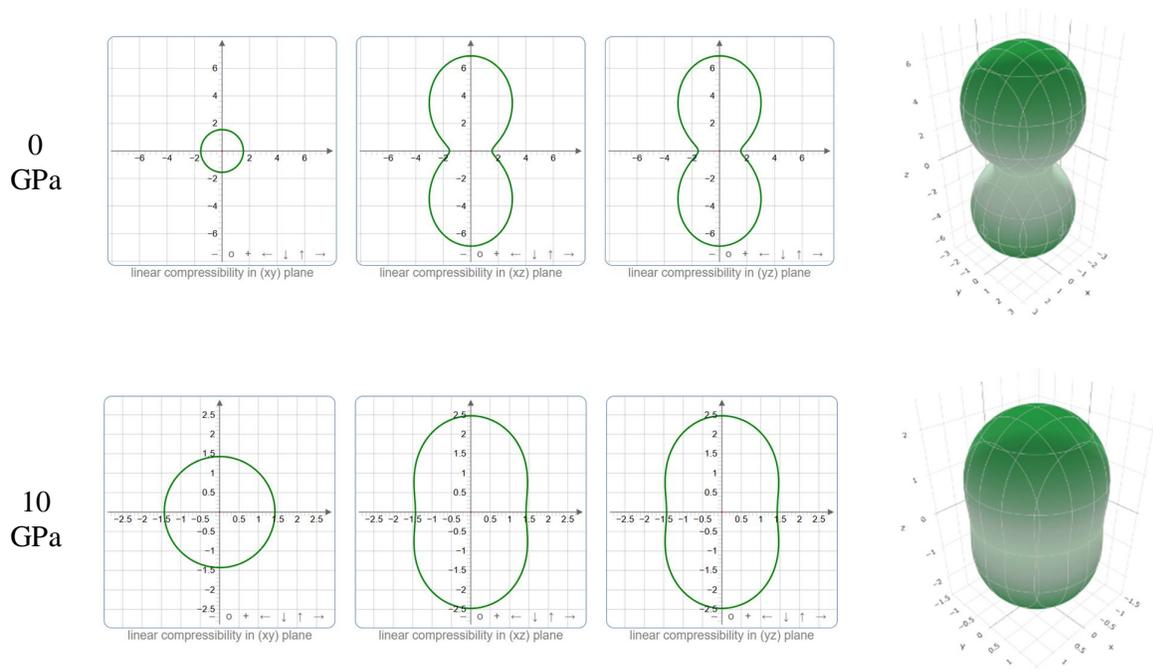

**Figure 6.** Directional variation in linear compressibility (*β*) of Ta$_2$Se compound.

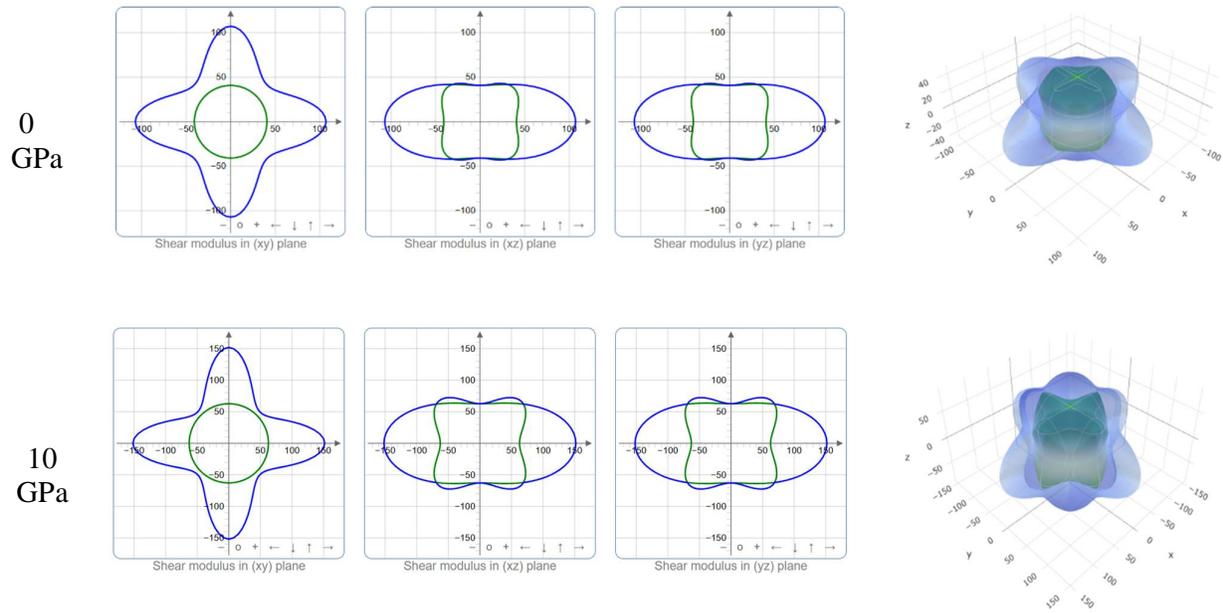

**Figure 7.** Directional variation in shear modulus (*G*) of Ta$_2$Se compound.

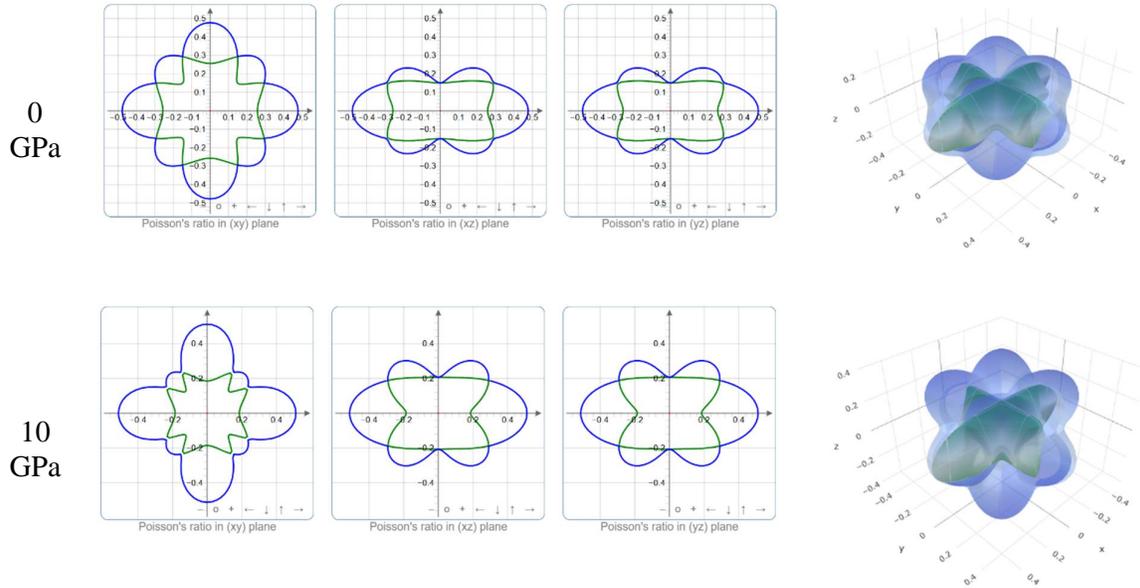

**Figure 8.** Directional variation in Poisson's ratio (*v*) of Ta$_2$Se compound.

## 3.4 Thermophysical properties

The parameters including sound velocities, the Debye temperature, melting temperature, heat capacity, minimum thermal conductivity, acoustic impedance and the Grüneisen parameter are essential for understanding a material's thermal behavior and predicting its potential applications. We calculated all these properties under varying hydrostatic pressures and summarized in **Table 6**.

**Table 6.** Pressure dependence of mass density ($\rho$ in gcm$^{-3}$), longitudinal, transverse, and mean sound velocities ($v_l$, $v_t$, and $v_m$ in ms$^{-1}$), Debye temperature ($\theta_D$ in K), melting temperature ($T_m$ in K), minimum thermal conductivity ($k_{min}$ in Wm$^{-1}$K$^{-1}$) and Grüneisen parameter ($\gamma_e$) of Ta$_2$Se.

| Pressure (GPa) | $\rho$ | $v_l$ | $v_t$ | $v_m$ | $\theta_D$ | $T_m$ | $k_{min}$ | $\gamma_e$ | Remarks |
|---|---|---|---|---|---|---|---|---|---|
| 0 | 13.09 | 3801.61 | 2096.05 | 2334.14 | 255.27 | 1285.65 | 2.61 | 1.66 | |
| 2 | 13.49 | 4005.12 | 2189.84 | 2440.13 | 269.94 | 1363.39 | 2.79 | 1.65 | |
| 4 | 13.81 | 4287.80 | 2306.99 | 2573.72 | 289.93 | 1536.61 | 2.99 | 1.75 | This work |
| 6 | 13.99 | 4456.10 | 2367.43 | 2643.50 | 299.77 | 1637.91 | 3.09 | 1.79 | |
| 8 | 14.16 | 4562.29 | 2408.64 | 2690.69 | 306.70 | 1701.84 | 3.17 | 1.81 | |
| 10 | 14.31 | 4614.16 | 2447.21 | 2732.90 | 312.71 | 1758.18 | 3.25 | 1.80 | |

**Debye temperature**

The Debye temperature, $\theta_D$ serves as a vital diagnostic tool for understanding a material's fundamental nature, effectively mapping the transition between low-temperature quantum behavior and high-temperature classical vibrations. It represents the threshold at which the entire spectrum of lattice vibrations becomes thermally accessible, culminating in a maximum frequency known as the Debye frequency, $v_D$. These two quantities are related by $hv_D = k_B\theta_D$, where $h$ is Planck's constant and $k_B$ is the Boltzmann constant. In general, materials with stiffer bonding exhibit higher Debye temperatures. We computed $\theta_D$ using Anderson's method [76],

$$\theta_D = \frac{h}{k_B}\left[\frac{3n}{4\pi}\left(\frac{N_A\rho}{M}\right)\right]^{\frac{1}{3}} v_m \qquad (16)$$

Here, $N_A$, $\rho$, $M$ and $n$ represent Avogadro's number, the crystal mass density, the molecular mass and the number of atoms per molecule, respectively. The average sound velocity $v_m$ in the solid is given by,

$$v_m = \left[\frac{1}{3}\left(\frac{2}{v_t^3} + \frac{1}{v_l^3}\right)\right]^{-\frac{1}{3}} \qquad (17)$$

Here, $v_t$ and $v_l$ denote the transverse and longitudinal sound velocities of the compound, which can be calculated from the elastic moduli and the density as follows,

$$v_l = \sqrt{\left(\frac{3B + 4G}{3\rho}\right)} \qquad (18)$$

And,

$$v_t = \sqrt{\frac{G}{\rho}} \qquad (19)$$

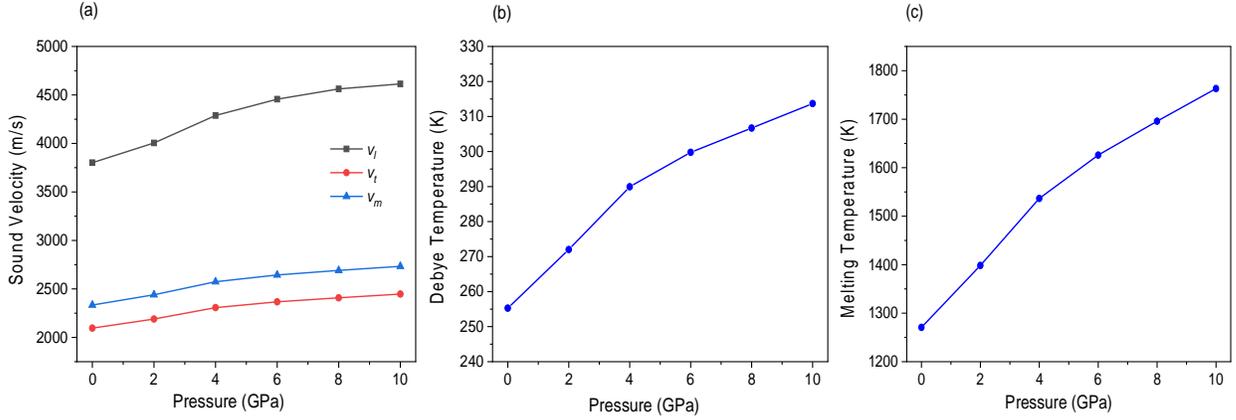

**Figure 9.** (a) Sound velocities Debye temperature and (b) Debye temperature and (c) Melting temperature of Ta$_2$Se under pressure.

Sound velocities (longitudinal $v_l$, transverse $v_t$, and average $v_m$) and Debye temperature ($\theta_D$) increase nearly steadily from 0 to 10 GPa [**Figure 9**(a) and (b)]. The longitudinal velocities exceed transverse ones, which is consistent with $C_{11}$ being larger than $C_{12}$ or $C_{44}$ for Ta$_2$Se. Here $\theta_D$ increases gradually with pressure. This is expected behavior, as applying pressure stiffens the crystal lattice and consequently raises the Debye temperature [77].

**Melting temperature**

The melting temperature $T_m$ defines the upper operating limit for high-temperature applications. A higher $T_m$ generally indicates stronger interatomic bonding and lower thermal expansion. $T_m$ was estimated from elastic constants using the following equation [78]:

$$T_m = \left[354 + 4.5\left(\frac{2C_{11}+C_{33}}{3}\right)\right]K \qquad (20)$$

Exploring how melting points change under extreme conditions is important in physics and materials science, so we also examine how pressure affects the melting temperature of our compound [79]. As depicted in **Figure 9**(c), the melting temperature exhibits a clear dependence on pressure. The melting temperature $T_m$ increases with pressure and stays high from 0 to 10 GPa, reflecting the pressure-induced stiffening of the material. High $T_m$ may arise from large heat of fusion, low entropy of fusion, or both [80]. The bonding strength of a crystal is strongly related to its overall structural stability $T_m$.

**Minimum thermal conductivity**

When the temperature exceeds the Debye temperature ($T > \Theta_D$), thermal conductivity approaches a lower bound, called the minimum thermal conductivity, $k_{min}$ (or $\Lambda_{min}$). It is important for high-temperature applications, as thermal conductivity generally decreases with rising temperature unitil it reaches a certain limit [81]. This minimum thermal conductivity ($k_{min}$) was computed using Clarke's model [82]:

$$k_{min} = k_B v_m \left(\frac{M}{n\rho N_A}\right)^{-\frac{2}{3}} \quad (21)$$

where, $k_B$ is the Boltzmann constant, $v_m$ is the average acoustic/sound velocity, $M$ is the molecular mass, $n$ is the number of atoms per unit cell, $\rho$ is the molecular weight and $N_A$ is the Avogadro constant. The estimated isotropic minimum thermal conductivity $k_{min}$ values are listed in the **Table 6** and show an increase with pressure, due to the rise in average sound velocity under compression.

**Grüneisen parameter**

The Grüneisen parameter ($\gamma$) is a dimensionless quantity that serves as a temperature dependent measure of anharmonicity, indicating the degree to which phonon vibrations in a crystal deviate from simple harmonic motion. There are five types of Grüneisen constants: elastic ($\gamma_e$), acoustic ($\gamma_a$), lattice ($\gamma_l$), electronic ($\gamma_{el}$) and thermodynamic ($\gamma_d$). For the structures studied, the elastic Grüneisen constant $\gamma_{el}$ was calculated using its relationship with the Poisson's ratio ($v$) [75, 83, 84]:

$$\gamma_e = \frac{3(1+v)}{2(2-3v)} \quad (22)$$

The value of $\gamma_e$ influences thermal expansion, thermal conductivity, elastic softening with temperature, and acoustic attenuation. The higher $\gamma_e$ is, the higher is the anharmonicity, the lower is the phonon thermal conductivity [85]. For Ta$_2$Se, $\gamma_e$ increases with pressure, indicating enhanced anharmonicity, consistent with trends in related thermophysical quantities. This reflects a similar trend seen in several other physical processes discussed earlier.

## 3.5 Bond Population Analysis

The bonding nature (ionic, covalent, or metallic) and effective valence were analyzed using Mulliken bond population analysis [40]. The Mulliken charge relates to a molecule's vibrational properties, measuring how its electronic structure shifts when atoms move. It also connects to properties like dipole moment, polarizability, electronic structure, and charge mobility in reactions [66] are also related to Mulliken charge. The spilling parameter and atomic charges obtained from these calculations are presented in **Table 7**.

**Table 7** shows that the charge spilling is very small (≈ 0.6% at both 0 and 10 GPa), so the basis set represents the valence charge of tetragonal Ta$_2$Se reliably. The Mulliken populations indicate that Ta atoms carry about ≈ 5.3-5.6 valence electrons, mainly of $d$ character, while Se has ≈ 5.1-5.3 electrons dominated by $p$ states. This confirms that the states near the Fermi level are mainly Ta-$d$/Se-$p$ hybridized.

**Table 7.** Charge spilling parameter (%), formal ionic charge, atomic Mulliken charges (electron), effective valence (electron) and Hirshfeld charge (electron) in Ta$_2$Se.

| Pressure (GPa) | Charge spilling | Species | Mulliken atomic populations | | | | FC | MC | EVC | HC | EVC | Ref. |
|---|---|---|---|---|---|---|---|---|---|---|---|---|
| | | | s | p | d | Total | | | | | | |
| 0 | 0.58 | Ta | 0.79 | 0.75 | 3.92 | 5.45 | +4 | -0.45 | 3.55 | 0.05 | 3.95 | This work |
| | | Ta | 0.74 | 0.68 | 3.84 | 5.26 | +4 | -0.26 | 3.74 | -0.03 | 3.97 | |
| | | Se | 1.07 | 4.21 | 0.00 | 5.28 | -2 | 0.72 | 1.28 | -0.02 | 1.98 | |
| 10 | 0.64 | Ta | 0.81 | 0.78 | 4.01 | 5.59 | +4 | -0.59 | 3.41 | 0.03 | 3.97 | |
| | | Ta | 0.78 | 0.68 | 3.86 | 5.32 | +4 | -0.32 | 3.68 | -0.03 | 3.97 | |
| | | Se | 0.91 | 4.18 | 0.00 | 5.09 | -2 | 0.91 | 1.09 | 0.00 | 2.00 | |

The formal ionic charges correspond to an ideal Ta$^{4+}$/Se$^{2-}$ picture, but the Mulliken (and Hirshfeld) charges are small and even have opposite sign (Ta slightly negative, Se slightly positive). This suggests that charge transfer is weak and the bonding is not purely ionic, but strongly covalent/metallic. Under 10 GPa, Ta becomes a bit more negative and Se more positive, showing a slight increase in Ta-Se charge transfer and hybridization, but the overall mixed covalent-metallic character of Ta$_2$Se remains.

**Table 8.** The calculated Mulliken bond overlap population of μ-type bond $P^\mu$, bond length $d^\mu$ (Å), and total number of bonds $N^\mu$, of Ta$_2$Se.

| Pressure (GPa) | Bond | $N^\mu$ | $P^\mu$ | $d^\mu$ | Remarks |
|---|---|---|---|---|---|
| 0 | Ta - Se | 2 | -0.17 | 2.689 | This work |
| | Ta – Se | 2 | -1.40 | 2.869 | |
| | Ta – Ta | 1 | 1.88 | 2.880 | |
| | Ta – Ta | 2 | 1.74 | 2.947 | |
| | Ta - Ta | 2 | 0.40 | 3.256 | |
| 10 | Ta - Se | 2 | 0.22 | 2.636 | |
| | Ta – Se | 2 | -1.95 | 2.840 | |
| | Ta – Se | 2 | -0.82 | 3.080 | |
| | Ta – Ta | 1 | 1.82 | 2.848 | |
| | Ta – Ta | 2 | 1.57 | 2.928 | |
| | Ta - Ta | 2 | 0.74 | 3.271 | |
| | Se - Se | 1 | -10.21 | 3.074 | |

**Table 8** uses the Mulliken bond-overlap population to quantify the strength and character of the bonds in tetragonal Ta$_2$Se. Positive overlap populations correspond to bonding states with electron density shared between the two atoms (covalent/metallic), whereas negative values indicate antibonding or strongly polar interactions (non-bonding). At both 0 and 10 GPa the Ta-Ta pairs have sizeable positive populations, so Ta atoms are linked by strong metallic-covalent bonds; their relatively short and pressure-contracting distances are consistent with a robust Ta metallic backbone. By contrast, most Ta-Se links show negative overlap, meaning that these interactions are dominated by antibonding character with only weak covalent sharing; only one Ta-Se contact becomes slightly bonding at 10 GPa (0.22 at 2.636 Å), indicating a small pressure-induced increase in Ta-Se covalency. The Se-Se pair at 10 GPa has a very large negative population (-10.21), which is strongly antibonding and confirms that no Se-Se covalent bond is formed at that distance. Overall, the data reveal a mixed bonding picture: a metallic-covalent Ta-Ta network combined with mostly ionic, partially antibonding Ta-Se interactions, and an essentially non-bonding (antibonding) Se-Se contact.

The results from both charge spilling and Mulliken bond overlap calculations provide a detailed picture of the bonding and charge distribution in Ta-Se under varying pressure conditions. The material exhibits a transition from ionic to more covalent behavior in Ta-Se bonds as pressure increases, while the metallic bonding in Ta-Ta interaction becomes more pronounced. New bonds emerge as pressure increases. Such pressure-driven electronic redistribution may influence transport, optical and magnetic responses of this material.

## 3.6   Electron Density Difference

Electron density difference (EDD) is a useful method to analyze the bonding characteristics in compounds, as it illustrates where charge builds up or is depleted around atoms. In covalent bonds, charge accumulation typically occurs between the bonded atoms. There is charge depletion around one atom and charge accumulation around the other in ionic bonding. Moreover, metallic bonding is demonstrated by uniform charge smearing. We have calculated the EDD of $Ta_2Se$ in ($\bar{1}10$) plane. **Figure 10** depicts the electron density difference of $Ta_2Se$ in the ($\bar{1}10$) plane under 0 GPa and 10 GPa hydrostatic pressures. The color scale in the center represents the electron density difference, ranging from -0.2267 to 0.1084 electrons/Å$^3$. Red and blue regions correspond to the highest negative and positive deviations in electron density respectively relative to the atoms.

At 0 GPa, [**Figure 10**(a)], the EDD map exhibits pronounced depletion lobes surrounding Ta sites, accompanied by accumulation both on Se sites and along the Ta-Se bond trajectories. This pattern signifies net electron transfer from Ta to Se together with a non-negligible build-up of density in the inter-nuclear region. The latter is a fingerprint of covalent sharing arising from Ta-5$d$/Se-4$p$ hybridization. Regions rendered in green correspond to electron density difference close to zero, i.e., areas where the bonded charge density closely resembles the superposition of isolated atoms.

Under 10 GPa, [**Figure 10**(b)], the blue accumulation around Se and along Ta-Se directions becomes more pronounced, while Ta-centered depletion intensifies slightly. This indicates that compression enhances Ta-Se hybridization and strengthens the covalent component of the Ta-Se bond, consistent with pressure-induced increases in Ta-Se overlap and charge transfer reported for other transition-metal chalcogenides [86].

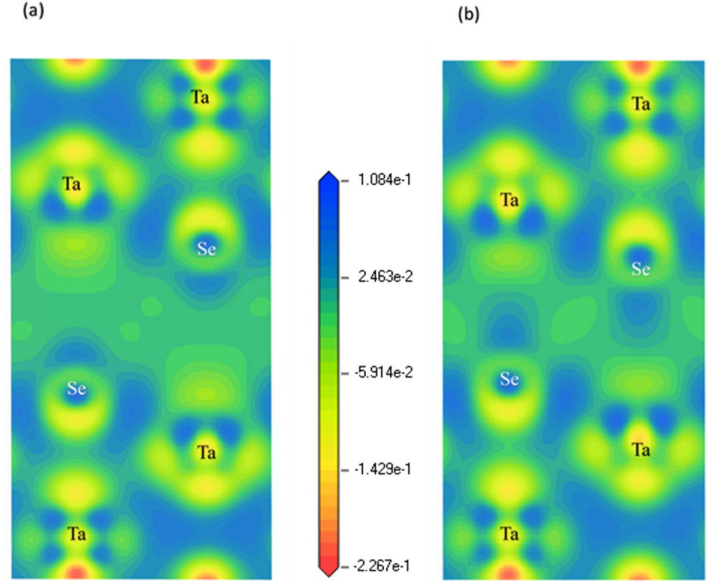

**Figure 10.** EDD maps of Ta$_2$Se on the ($\bar{1}$10) plane at (a) 0 GPa and (b) 10 GPa. Blue (positive) denotes charge accumulation and red (negative) denotes depletion relative to a superposition of neutral atoms; the central color bar quantifies charge difference in units of electronic charge. The maps show Ta-centered depletion and Se/bond-centered accumulation, evidencing Ta to Se charge transfer with appreciable covalency that strengthens under pressure.

### 3.7 Optical Properties

In optical properties, we study how a material response to the irradiation of electromagnetic energy. Optical response was evaluated from frequency-dependent dielectric and related functions governed by the electronic structure. Because Ta$_2$Se is metallic, a Drude contribution was included using a plasma energy of 5 eV, damping of 0.01 eV, and 0.5 eV smearing at 0 and 10 GPa [87, 88]. Since Ta$_2$Se is elastically anisotropic, optical parameters are expected to show direction dependence. To study anisotropy, spectra were computed for electric field along [100] and [001] polarization as a function of photon energy up to 30 eV for 0 GPa and 10 GPa pressures [**Figure 11**(a-f)]. The macroscopic response is described by the complex dielectric function,

$$\varepsilon(\omega) = \varepsilon_1(\omega) + i\varepsilon_2(\omega) \tag{23}$$

The frequency dependent real part, $\varepsilon_1(\omega)$ reflects polarizability and phase velocity, whereas the imaginar part, $\varepsilon_2(\omega)$ describes absorption linked to interband transitions and the band structure [80]. In metals, intraband (Drude) transitions dominate at low energy, while interband transitions shape higher-energy features. Large negative $\varepsilon_1(\omega)$ at low energy indicates Drude-like metallic response [**Figure 11**(a)]. $\varepsilon_2(\omega)$ approaches zero near ~23 eV, implying transparency above this energy. The effective plasma frequency $\omega_p$ lies near ~23 eV for both polarizations and at both pressures.

Optical absorption starts at zero photon energy for both pressures, confirming the metallic nature of Ta$_2$Se. This also provides insight into the material's potential application for efficient solar energy conversion. **Figure 11**(b) shows its absorption spectra. High absorption occurs mainly in the ~ 5 to 20 eV range. For both pressures, absorption coefficient $\alpha(\omega)$ is slightly higher along the [001] polarization direction compared to [100]. α(ω) decreases sharply at ~ 22 eV for both pressures, respectively, in alignment with the position of the loss peaks.

Optical Reflectivity R(ω) is highest at zero photon energy at both 0 and 10 GPa and decreases nearly linearly up to ~2 eV, reflecting the free-electron response of a metal **Figure 11**(c). $R_{max}$ at zero photon energy indicates that the compound exhibits a shiny, metallic-like appearance. R remains above 50% up to ~22 eV, indicating strong solar reflectance, while pressure modifies R through changes in the electronic structure and DOS at $E_F$.

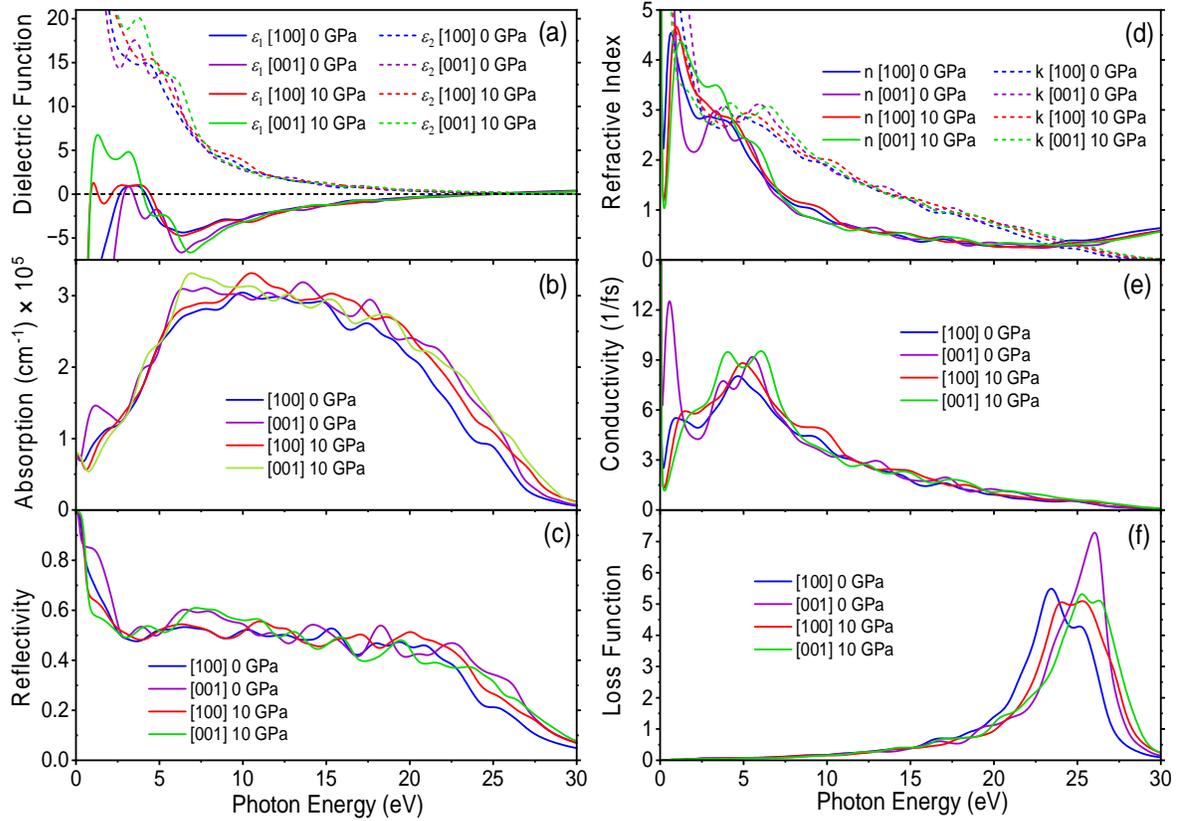

**Figure 11.** The energy (or, equivalently, frequency) dependent (a) dielectric function (b) absorption coefficient (c) reflectivity (d) refractive index (e) optical conductivity and (f) loss function of $Ta_2Se$ with electric field polarization vectors along [100] and [001] directions at 0 GPa and 10 GPa pressure.

The refractive index as a complex function of frequency, expressed as N(ω) = n(ω) + ik(ω), where k(ω) is the extinction coefficient. The real part, n(ω), reflects the wave's speed in the material, while k(ω) shows how much light is absorbed. High k(ω) means stronger light attenuation, important for designing optoelectronic devices. As shown in **Figure 6**(d), n(ω) peaks at zero energy and decreases with photon energy. The compound has high refractive indices in the visible range under both pressures, making it suitable for optoelectronic displays.

The optical conductivity, σ(ω) describes the dynamic response of mobile carriers to incident photons. For $Ta_2Se$, σ(ω) begins at zero photon energy, indicating the absence of a band gap, a finding that aligns perfectly with our calculated band structure [**Figure 12**] and TDOS results [**Figure 15**]. The compound's photoconductivity increases with photon energy, reaches its maximum, progressively decreases with further energy, and goes to zero at about 30 eV. The spectral features generally follow ε₂(ω), with peaks arising from bulk plasmon excitations and transitions of electrons from the valence to the conduction band [89].

**Figure 11**(f) displays the electron energy-loss function, L(ω), for the studied materials at 0 and 10 GPa. L(ω) represents the energy lost by fast electrons through processes like plasmon and phonon excitations, interband and intraband transitions, inner-shell ionizations, and Cerenkov radiation [89]. The highest peak in energy-loss function spectrum for a material, associated with the plasma resonance and occur where $\varepsilon_2 < 1$ and $\varepsilon_1 = 0$ [90]. The highest peak of L(ω) for [100] and [001] at 0 GPa and 10 GPa is about 23 eV, 27 eV and 23.5 eV and 25 eV, respectively. A material becomes transparent and exhibits insulator-like optical properties at photon frequencies above its plasma frequency, $\omega_p$. The peak energy also marks the transition from metallic to dielectric behavior. According to the Drude model, the plasma frequency is given by:

$$\omega_p = \sqrt{\frac{n_e e^2}{\varepsilon_0 m^*}} \tag{24}$$

Where, e is the elementary charge, $n_e$ is the electron density, $m^*$ is the effective mass of the electrons and $\varepsilon_0$ is the permittivity of free space. The basic feature of the connection between free electron density and collective electronic excitations in metals and other conductive materials is this relationship [87, 88].

## 3.8    Electronic Properties

**Electronic Band Structure**

Many functional properties of a solid are set by valence and conduction electrons. Their response is governed by the energy-momentum relation E(*k*) across the Brillouin zone, i.e., the electronic band structure. The band structure is crucial for understanding a material's properties, including electrical thermal conductivity, the Hall effect, heat capacity, magnetic and optoelectronic behavior. **Figure 12** shows the electronic band structure of tetragonal Ta₂Se along the high-symmetry path Γ-X-M-Γ-Z-R-A-Z under hydrostatic pressures from 0 to 10 GPa, with the Fermi level, $E_F$, set to zero energy.

With increasing pressure (colour sequence from blue to red), no band gap opens at any *k*-point; Ta₂Se remains metallic up to at least 10 GPa. Compression mainly enhances band dispersion, resulting in increasing curvature for bands near $E_F$ both in-plane (Γ-X-M) and out-of-plane (Γ-Z-R-A-Z). This trend reflects stronger Ta-5*d*/Se-4*p* overlap at shorter distances and matches pressure-driven band broadening reported for related Ta-chalcogenides and layered dichalcogenides [91]. Pressure also shifts several band extrema relative to $E_F$. Near M and along M-Γ and Γ-Z, shallow minima just below $E_F$ move upward, while nearby conduction features bend downward, increasing electron-hole overlap and likely creating additional Fermi-surface sheets at higher pressure. This pattern is consistent with an emerging Lifshitz-type electronic topological transition, where extrema approach or cross $E_F$ without changing crystal symmetry [92].

The pressure dependence of electronic degeneracies at high-symmetry points also deserves attention. At ambient pressure, several bands are nearly degenerate near Γ, Z and R, but compression lifts these degeneracies and drives band splitting. This is compatible with anisotropic lattice compression, which alters the Ta crystal field and shifts the balance between Ta-Ta metallic bonding and Ta-Se covalency; similar behavior often appears before structural or electronic transitions in layered-tetragonal chalcogenides [93].

The highly dispersive bands along Γ-X, M-Γ, Z-R, and A-Z indicate electronic character at all pressures, with some crossing the Fermi level, suggesting they dominate charge transport. In contrast, the Γ-Z bands are nearly flat at all pressures. Broadly dispersing bands imply low effective mass and high charge carrier mobility [94-96]. The material exhibits pronounced anisotropy in charge transport both in momentum space and along different crystallographic directions. This is indicated by the nearly flat E($k$) bands at the Γ point, which contributes to van Hove singularities in the Γ-Z region.

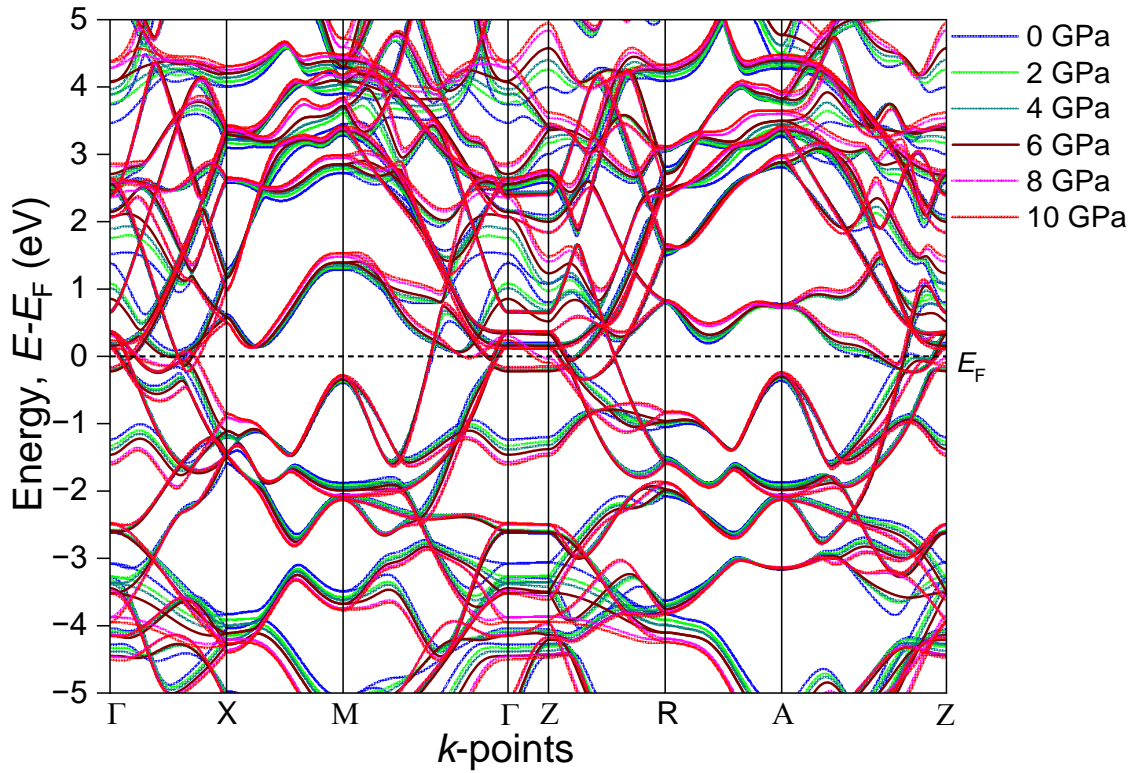

**Figure 12.** Comparison of electronic band structures of Ta$_2$Se mapped along high-symmetry paths within the first Brillouin zone, across a pressure range of 0 to 10 GPa in increments of 2 GPa.

**Figure 13** highlights the strong effect of spin-orbit coupling (SOC) on Ta$_2$Se's band structure. SOC lifts band degeneracy and shifts energies in both valence and conduction bands, especially near the Fermi level. The flat bands at the Γ point create van Hove singularities in the density of states, which are enhanced by SOC and may play a key role in the material's superconducting behavior which is shown in **Figure 13**(a) and (b) for two different pressures 0 GPa and 10 GPa respectively.

22 | P a g e

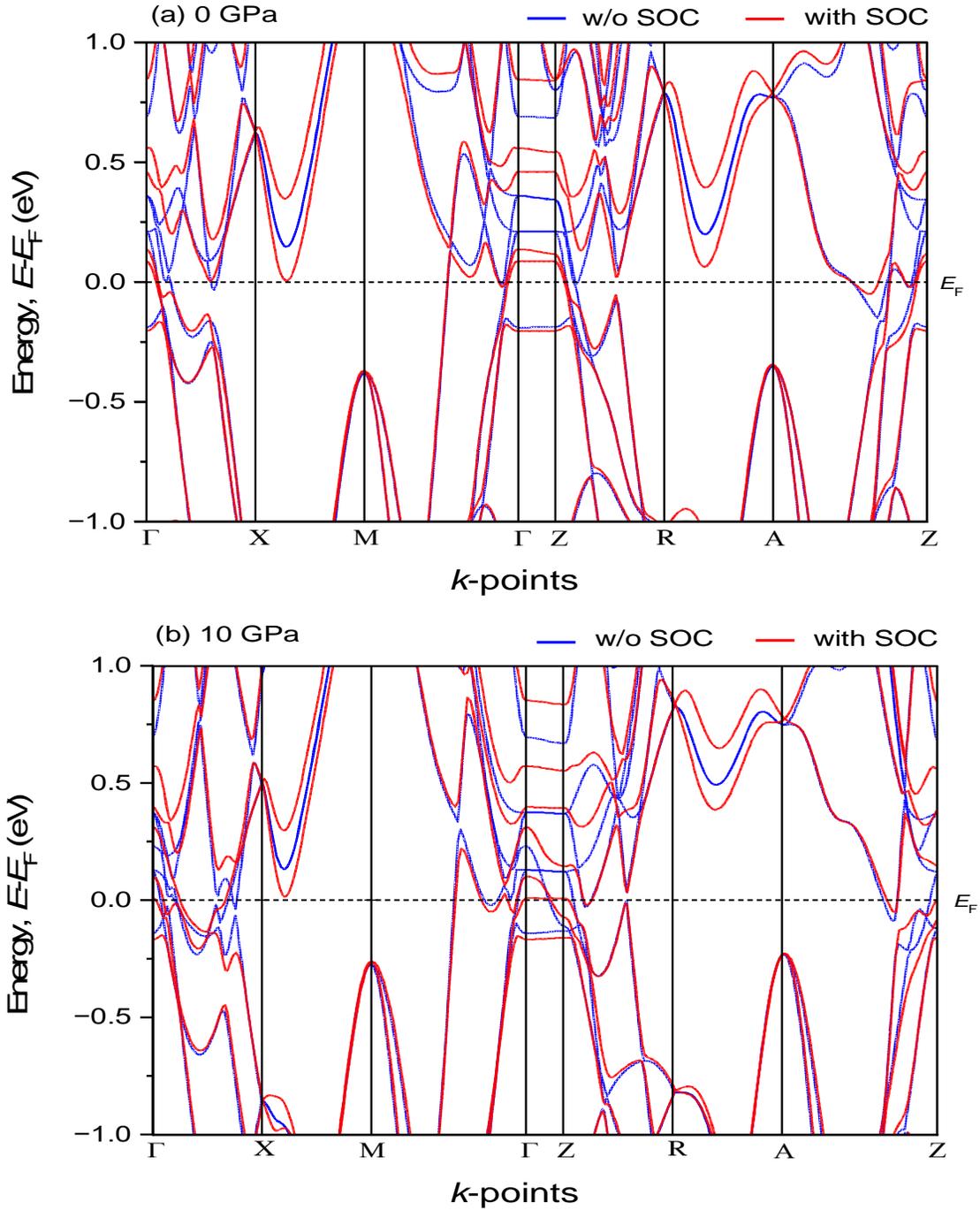

**Figure 13.** Band structures of Ta$_2$Se along Γ-X-M-Γ-Z-R-A-Z within ±1 eV of $E_F$ at (a) 0 GPa and (b) 10 GPa. Blue solid: without SOC. Red solid: with SOC. SOC lifts near-degeneracies and produces avoided crossings. Pressure increases dispersion (bandwidth) and slightly shifts SOC-split features, subtly modifying the Fermi pockets near $E_F$.

The projected (fatband) band structures of Ta$_2$Se, with and without spin-orbit coupling (SOC), are shown in **Figure 14** at ambient pressure (0 GPa). In both cases, the Fermi level is predominantly dominated by Ta $d$-orbital electrons, highlighting the key role of metal-metal bonding in the material's stability and superconductivity, while the *p*-orbitals from Se atoms mainly contribute below -2 eV to stabilize the structure, in agreement with earlier calculations and photoemission studies on metal-rich Ta-Se compounds [12]. This behavior is consistent with the layered tetragonal P4/*nmm* structure, where edge-sharing TaX$_4$ (X = chalcogen) units form a three-dimensionally connected Ta-Ta network that gives rise to wide *d*-bands at the Fermi level [97]. There is a bandgap opening occurs by including SOC which may introduce topological features [98].

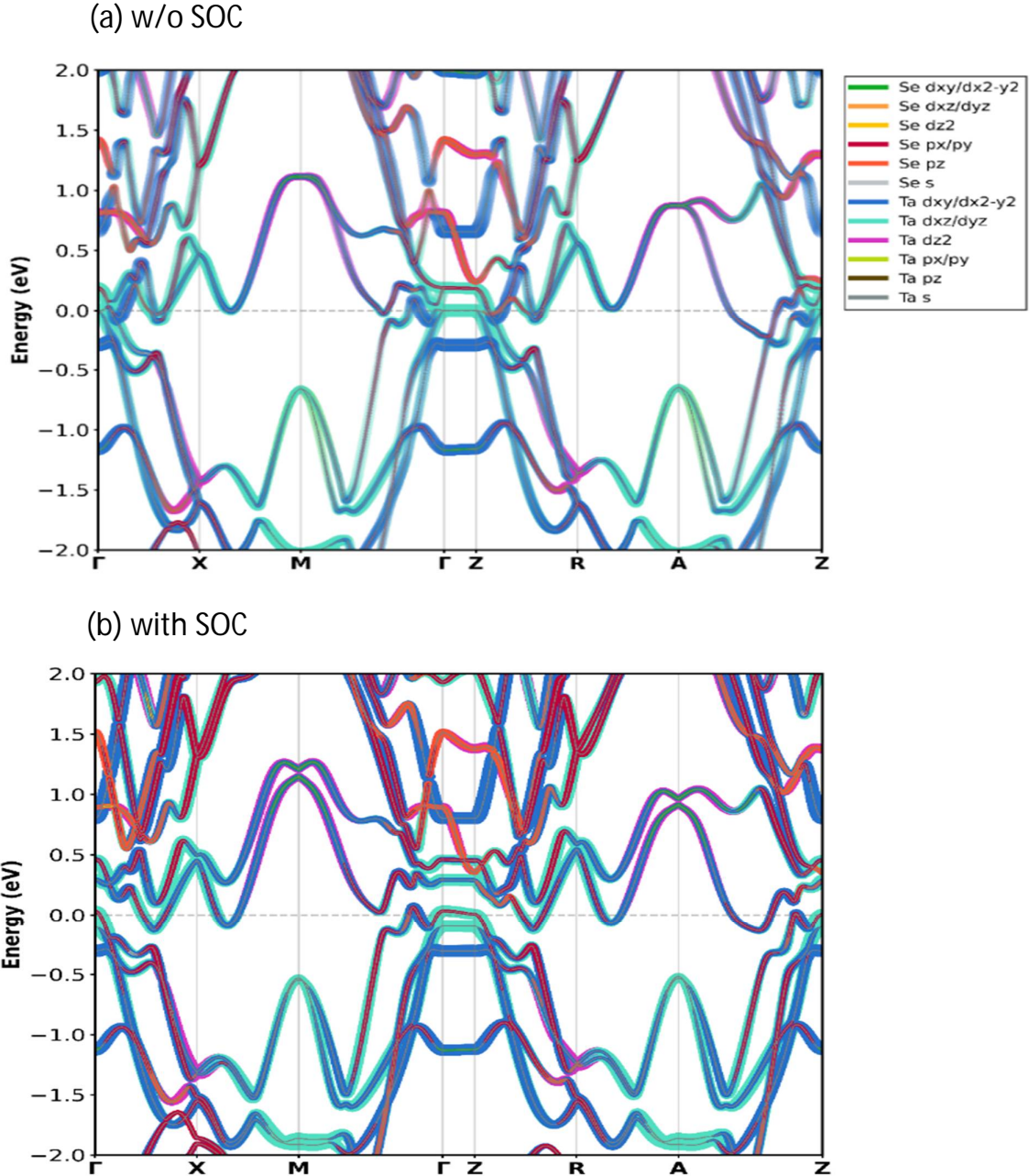

**Figure 14.** Orbital projected electronic band structures of Ta$_2$Se along Γ-X-M-Γ-Z-R-A-Z within ± 2 eV of $E_F$ for (a) w/o SOC and (b) with SOC at ambient pressure (0 GPa).

**Electronic Energy Density of States (EDOS)**

**Figure 16** shows the calculated total and partial density of states (TDOS and PDOS) of $Ta_2Se$ at 0 GPa and 10 GPa. Finite TDOS at $E_F$, indicating metallic conductivity at both pressures. Near $E_F$, the dominant weight comes from Ta-$5d$ and Se-$4p$ states, and their strong overlap signals pronounced hybridization, consistent with strong covalent bonding.

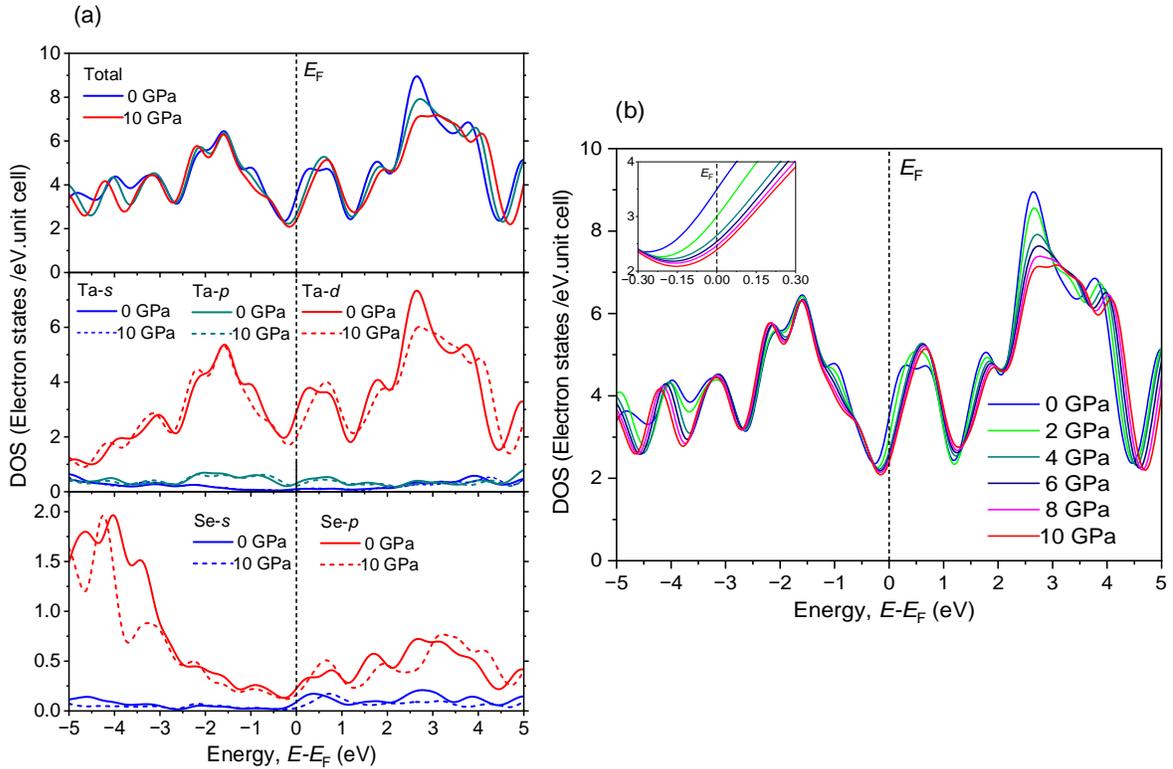

**Figure 15.** Pressure dependent electronic density of states (DOS) of $Ta_2Se$. (a) Total DOS and orbital-projected DOS at 0 GPa and 10 GPa, showing contributions from Ta-$s$, Ta-$p$, Ta-$d$ and Se-$s$, Se-$p$ states. (b) Evolution of the total DOS under hydrostatic pressure from 0 to 10 GPa (step 2 GPa); the inset highlights the DOS variation near Fermi level ($E_F$). Energies are plotted relative to $E_F$ (vertical dashed line at ($E$- $E_F$ = 0 eV)

The location of the Fermi level and the TDOS value at the Fermi energy, N($E_F$), are related to a compound's electronic stability [99, 100]. A pseudogap or quasigap in TDOS near $E_F$ can separate bonding states from nonbonding or antibonding states and is widely associated with stability states [101, 102]. At all pressures, the Fermi levels lie near the pseudogap minima [**Figure 15**]. Literature [101] cites two main causes for pseudogaps: ionic effects and orbital hybridization. The pseudogap near the Fermi level also suggests the presence of directional bonding [103]. This promotes covalent bond formation and improves the material's mechanical strength. In particular, near the Fermi level, strong bonding mainly comes from hybridization between Ta-$5d$ and Se-$4p$ states, leading to directional covalent bonding. This is consistent with the electronic charge density maps and Mulliken bond population analysis. A Fermi level within the pseudogap typically indicates a well-ordered, stable compound with a high melting point. If it shifts into the antibonding region beyond the pseudogap, the material becomes less stable and may tend toward a disordered or glassy state [104]. For $Ta_2Se$, $E_F$ is reported to fall in the antibonding side [**Figure 15**]. The effect of pressure on TDOS at $E_F$ has shown in **Table 9** and **Figure 15**(b).

**Table 9.** Calculated total density of states at the Fermi level, (N($E_F$) in electrons/eV-unit cell) of Ta$_2$Se under different pressures.

| Pressure (GPa) | N($E_F$) | Ref. |
|---|---|---|
| 0 | 3.494 | |
| 2 | 3.005 | |
| 4 | 2.655 | [This Work] |
| 6 | 2.540 | |
| 8 | 2.462 | |
| 10 | 2.389 | |

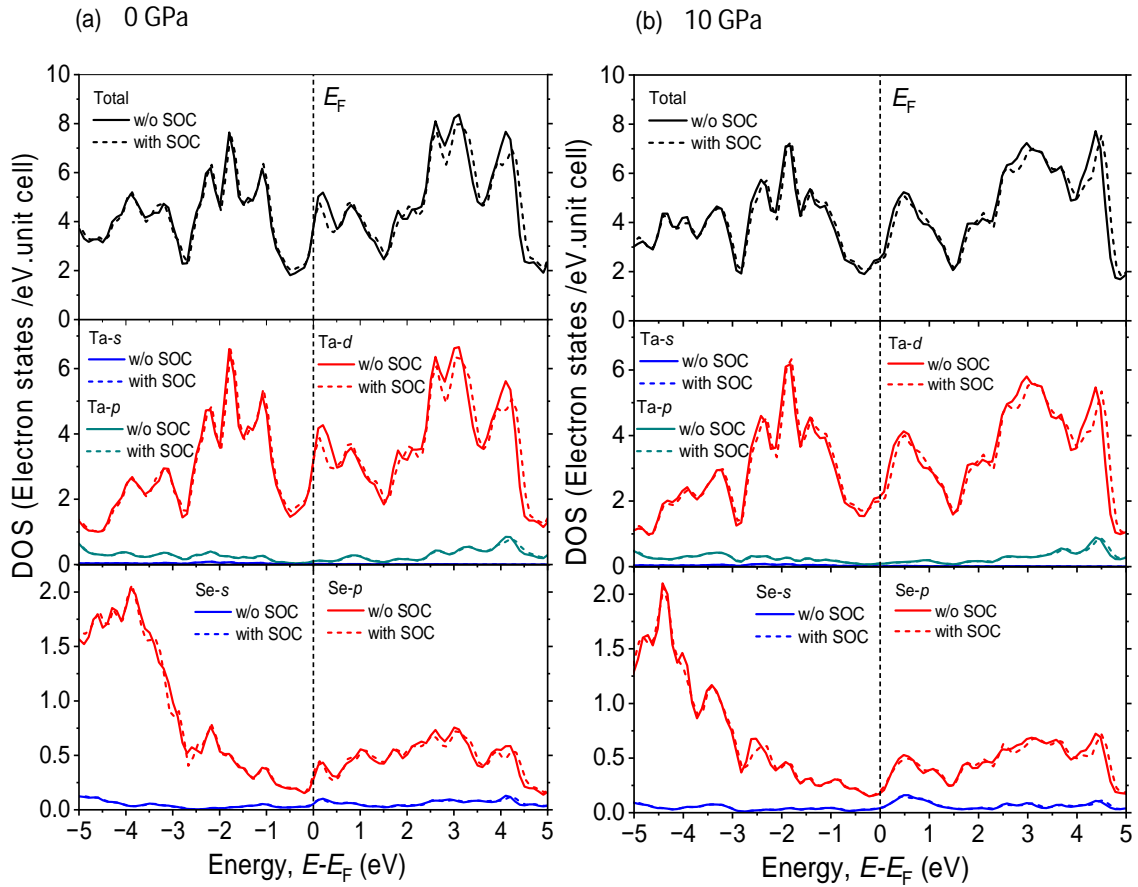

**Figure 16.** Spin-orbit coupling (SOC) effect on the electronic density of states (DOS) of Ta$_2$Se at different pressures: (a) 0 GPa and (10) GPa. Total DOS (top panels) and orbital-projected DOS (middle/bottom panels) are compared without SOC and with SOC, showing contributions from Ta-*s*, Ta-*p*, Ta-*d* and Se-*s*, Se-*p* states. The Fermi level is set to 0 eV (vertical dashed line) and energies are plotted relative to $E_F$ (vertical dashed line at ($E - E_F = 0$ eV).

Analyzing the **Table 9**, it is evident that as pressure increases from 0 GPa to 10 GPa, there is a gradual decrease in the total density of states N($E_F$) at the Fermi level. For instance, N($E_F$) decreases from 3.494 at 0 GPa to 2.389 at 10 GPa. This reduction suggests that the number of electronic states available at the Fermi level diminishes as the system is subjected to higher pressure, which can be associated with pressure-induced modifications in the electronic structure, such as band broadening or shifting of energy bands. Furthermore, the Coulomb pseudopotential $\mu^*$ [105] would also show a slight decrease. This subtle variation in $\mu^*$ could indicate a minor reduction in the electron-electron repulsion with increasing pressure, which might affect the superconducting properties of the material. Overall, these pressure-dependent changes in N($E_F$) and $\mu^*$ reflect the complex interplay between electronic structure and external pressure [106] in Ta$_2$Se, suggesting possible implications for its conductivity under different pressure conditions. The SOC effect on TDOS and PDOS has shown in **Figure 16**(a) and (b) for different pressures 0 GPa and 10 GPa respectively.

**Fermi Surface**

The Fermi surface (FS) in momentum space separates occupied from unoccupied electronic states at T = 0 K and provides a compact description of many electrical, magnetic, thermal and optical properties. Carrier dynamics depend on FS shape and sheet topology. Strong curvature is typically linked to smaller effective mass (m*), while flatter regions correspond to heavier carriers. Anisotropic FS geometries lead to direction-dependent Fermi velocity and hence anisotropic transport [107, 108]. Electrons near the Fermi surface play a key role in forming the superconducting state [109]. The electrical properties of metallic systems can often be inferred from the shape of the Fermi surface within the Brillouin zone. **Figure 17** presents the FS sheets of the three bands crossing $E_F$ at 0 and 10 GPa. Ta$_2$Se shows one electron-like and two hole-like sheets extending along Γ-Z, with an approximately symmetric appearance in top view. The corresponding crossings in the band structure are color-marked, and the bands intersect $E_F$ multiple times across the high-symmetry BZ region.

We show FS discrete viewed from the top for each corresponding band in **Figure 18**. The 69th band in **Figure 18**(a) consists of cylindrical like shape or hole-like sheets enclosing the middle of the Brillouin zone keeping the Γ-Z direction as its axis. As we apply 10 GPa pressure, the Fermi surface shrinks, and the pocket's shape undergoes some deformation.

At 0 GPa, band 70 exhibits a complex Fermi surface with multiple pockets around key symmetry points, suggesting localized or directionally dispersed electronic states. At elevated pressure of 10 GPa, these pockets change shape or merge, indicating a shift in the material's electronic structure. This could impact electron transport by modifying available conduction pathways.

In band 71, the Fermi surface at 0 GPa is compact, indicating more localized electron states. With 10 GPa pressure, it shrinks further, suggesting a reduction in density of states near the Fermi level. This could make the material more metallic and alter its conductivity by reducing electronic scattering.

The FS of band 69 and band 70 [**Figure 18**(a) and (b)] are classified as hole-like. The FS of band 71 [**Figure 18**(c)] is known as electron-like [110]. The pressure effect on the FS is significant, so we performed this calculation for both 0 GPa and 10 GPa.

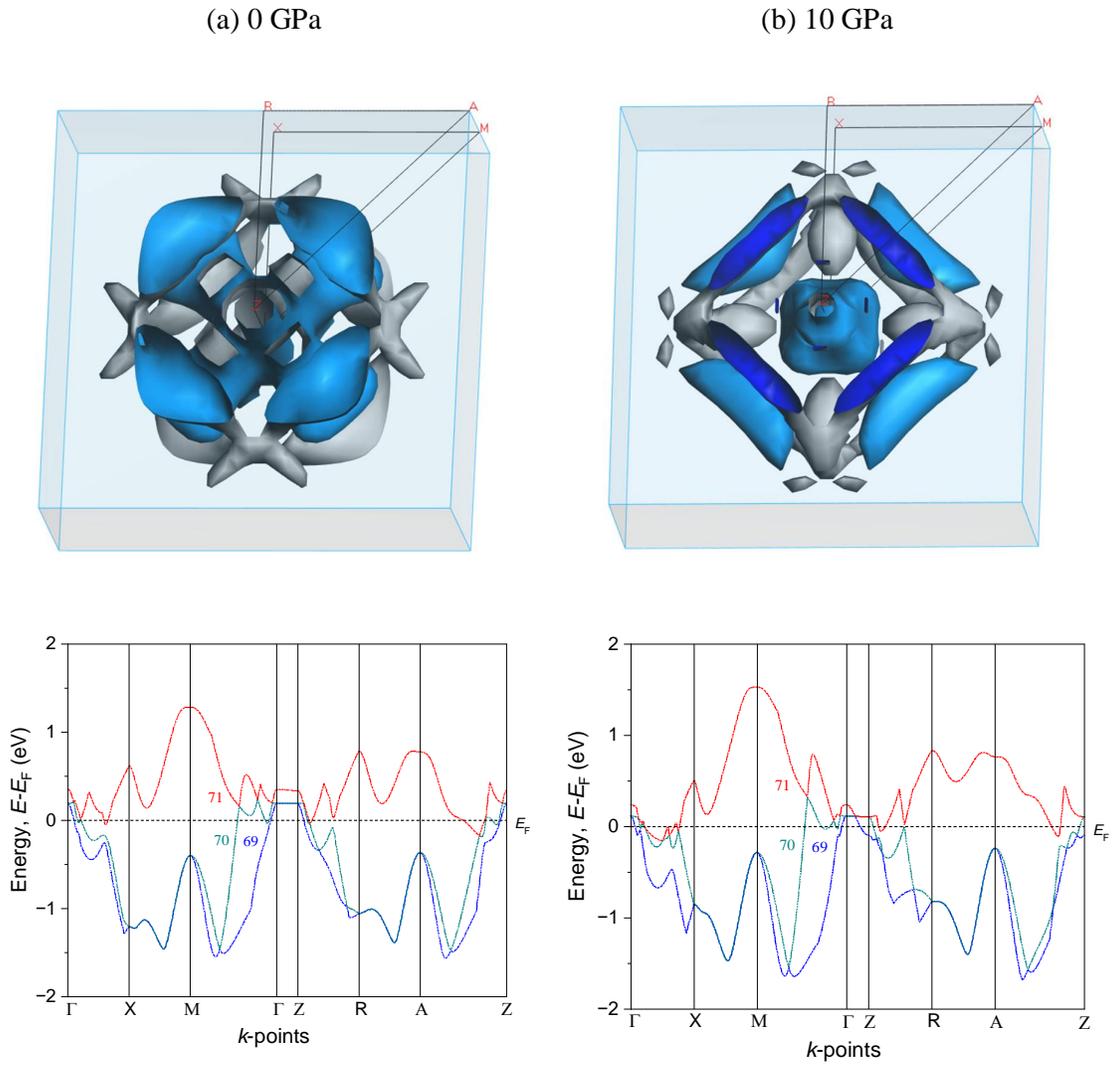

**Figure 17.** (a) and (b) Calculated three bands crossing Fermi surface of Ta$_2$Se for 0 GPa and 10 GPa respectively, (c) and (d) Associated bands (69, 70, 71) cross the Fermi level in the band structure.

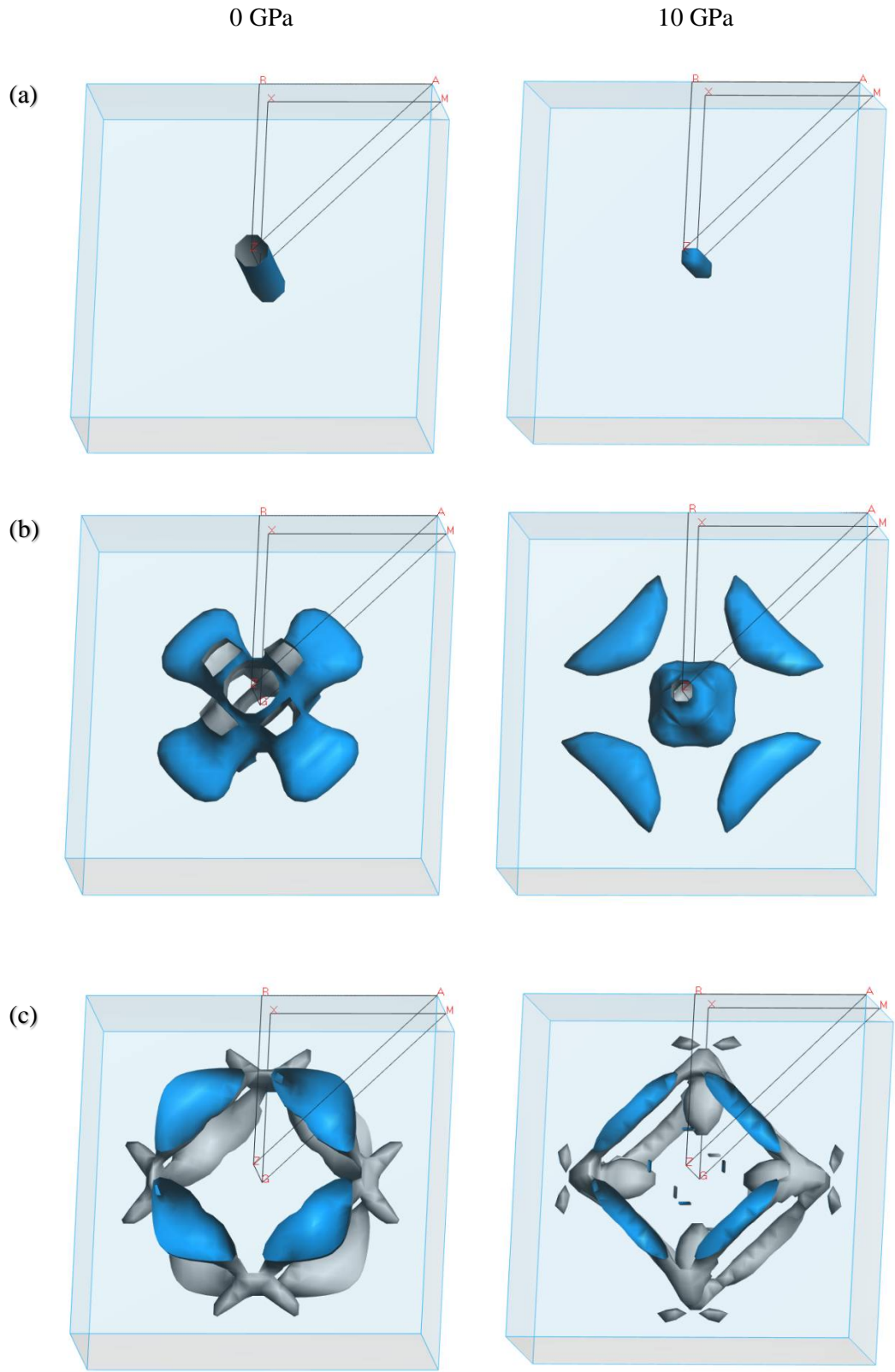

**Figure 18.** The fermi surfaces of Ta$_2$Se for bands (a) 69, (b) 70 and (c) 71 respectively for two different pressure 0 GPa and 10 GPa.

## 3.9 Phonon Dynamics

Phonon dispersion and density of states (PhDOS) affect many dynamical and thermodynamic properties, such as lattice stability, phase transitions, heat transport, Helmholtz free energy, thermal expansion and heat capacity. Since electron-phonon coupling (EPC) depends on the phonon spectrum, the PhDOS is also important for superconductivity and optical behaviors influenced by optical phonons. These properties are studied using DFPT and the finite-displacement method [111, 112], phonon dispersions and total PhDOS were computed along high-symmetry directions at 0 and 10 GPa [**Figure 19**]. Both pressures show no imaginary frequency across the Brillouin zone, indicating dynamical stability in each case.

A unit cell with N atoms has 3N phonon branches, consisting of 3 acoustic and (3N-3) optical modes. The acoustic modes (two transverse and one longitudinal) are related to sound propagation and elastic stiffness. Optical modes arise from out-of-phase atomic vibrations, where one atom moves one way while its neighbor moves the opposite. For $Ta_2Se$, the unit cell has 6 atoms, producing 18 phonon modes: 3 acoustic (red) and 15 optical, with the optical modes having nonzero frequencies at the $\Gamma$ point.

Mode resolved PhDOS indicates that Ta (the heavier atom) mainly contributes to the acoustic and lower optical vibrations at both pressures, whereas Se dominates the upper optical region. The strongest PhDOS peaks appear near 4.28 THz (0 GPa) and 4.23 THz (10 GPa). There is a little frequency gap between two optical branches at 10 GPa. At both pressures, the low-energy optical modes merge with the acoustic modes, indicating the absence of a phonon band gap between them. This suggests that the compound is favorable for efficient thermal transport. The highest optical frequency occurs near M, decreasing from about 8.88 THz (0 GPa) to 8.48 THz (10 GPa).

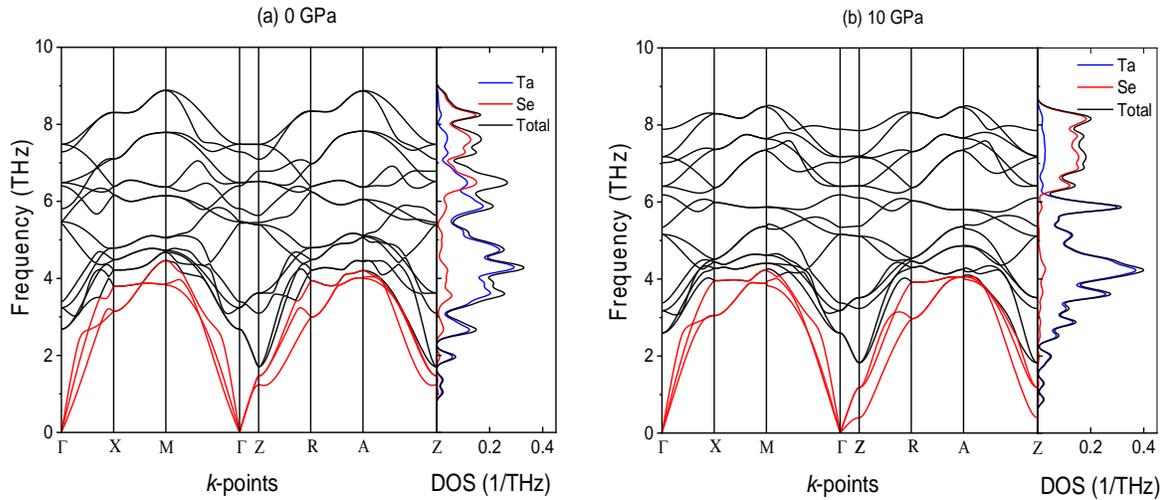

**Figure 19.** Phonon dispersion spectra and corresponding phonon density of states (DOS) at (a) 0 GPa and (b) 10 GPa in the ground state.

According to the group theory [113], $Ta_2Se$ has a total of 18 vibrational modes. The modes at the $\Gamma$ point are classified as follows:

$$\Gamma = (A_{2u} + E_u)_{acoustic} + 3A_{1g} + 3E_g + 2A_{2u} + 2E_u \qquad (25)$$

This result corresponds to 3 acoustic and 15 optical branches at Γ, including doubly degenerate $E_g$ and $E_u$ modes. Among the optical modes, $3A_{1g}$ and $3E_g$ are Raman-active, while $2A_{2u}$ and $2E_u$ are IR-active [114]. Our calculations confirmed the presence of six Raman-active modes: three $A_{1g}$ modes located at 101.5, 190.4, and 216.4 cm$^{-1}$, and three $E_g$ modes at 45.9, 161.1 and 185.3 cm$^{-1}$. Additionally, we identified four IR-active modes, consisting of two $E_u$ modes at 89.7 cm$^{-1}$ and 166.2 cm$^{-1}$ and two $A_{2u}$ modes at 169.9 cm$^{-1}$ and 190.2 cm$^{-1}$. The analysis also revealed two acoustic phonon modes, classified as $A_{2u}$ and $E_u$ modes. Our calculated results show good agreement with the existing data [13].

### 3.10 Superconductivity

The superconducting properties of Ta$_2$Se are evaluated using electron-phonon coupling (EPC) calculations. The superconducting transition temperature, $T_c$, is estimated using the Allen-Dynes-modified McMillan formula [115, 116]:

$$T_c = \frac{\omega_{log}}{1.2} exp\left[\frac{-1.04(1+\lambda)}{\lambda - \mu^*(1+0.62\lambda)}\right] \qquad (26)$$

Where, $\omega_{log}$ is the logarithmically averaged characteristic phonon frequency, $\mu^*$ is the effective Coulomb pseudopotential. The integrated EPC constant can be evaluated by

$$\lambda(\omega) = 2\int_0^\infty \frac{\alpha^2 F(\omega)}{\omega} d\omega \qquad (27)$$

The Eliashberg electron-phonon spectral function, $\alpha^2F(\omega)$ is written as follows [115]:

$$\alpha^2 F(\omega) = \frac{1}{2\pi N(E_F)} \sum_{qv} \frac{\gamma_{qv}}{\hbar\omega_{qv}} \delta(\omega - \omega_{qv}) \qquad (28)$$

where, $\omega_{qv}$ is the phonon frequency and $\gamma_{qv}$ is the phonon linewidth for the phonon with wave vector $q$ and branch $v$.

$$\gamma_{qv} = 2\pi\omega_{qv} \sum_{i,j} \int_{BZ} \frac{d^3k}{\Omega_{BZ}} |g_{qv}(k,i,j)|^2 \delta(\varepsilon_{k,i} - \varepsilon_F)\delta(\varepsilon_{k+q,j} - \varepsilon_F) \qquad (29)$$

Here, $g_{qv}(k, i, j)$ is the matrix of the EPC, and $\varepsilon_{q,i}$ is the electronic energy.

**Figure 20**(a) shows the phonon dispersions weighted by the magnitude of EPC. To determine which phonon mode drive superconductivity in this compound, we compare the atom-resolved phonon dispersions with where the EPC is strongest and how it is distributed across the branches. For Ta$_2$Se, the softened modes around the Γ and Z points show obvious EPC. These contributions are dominated by in-plane Ta-xy vibrations.

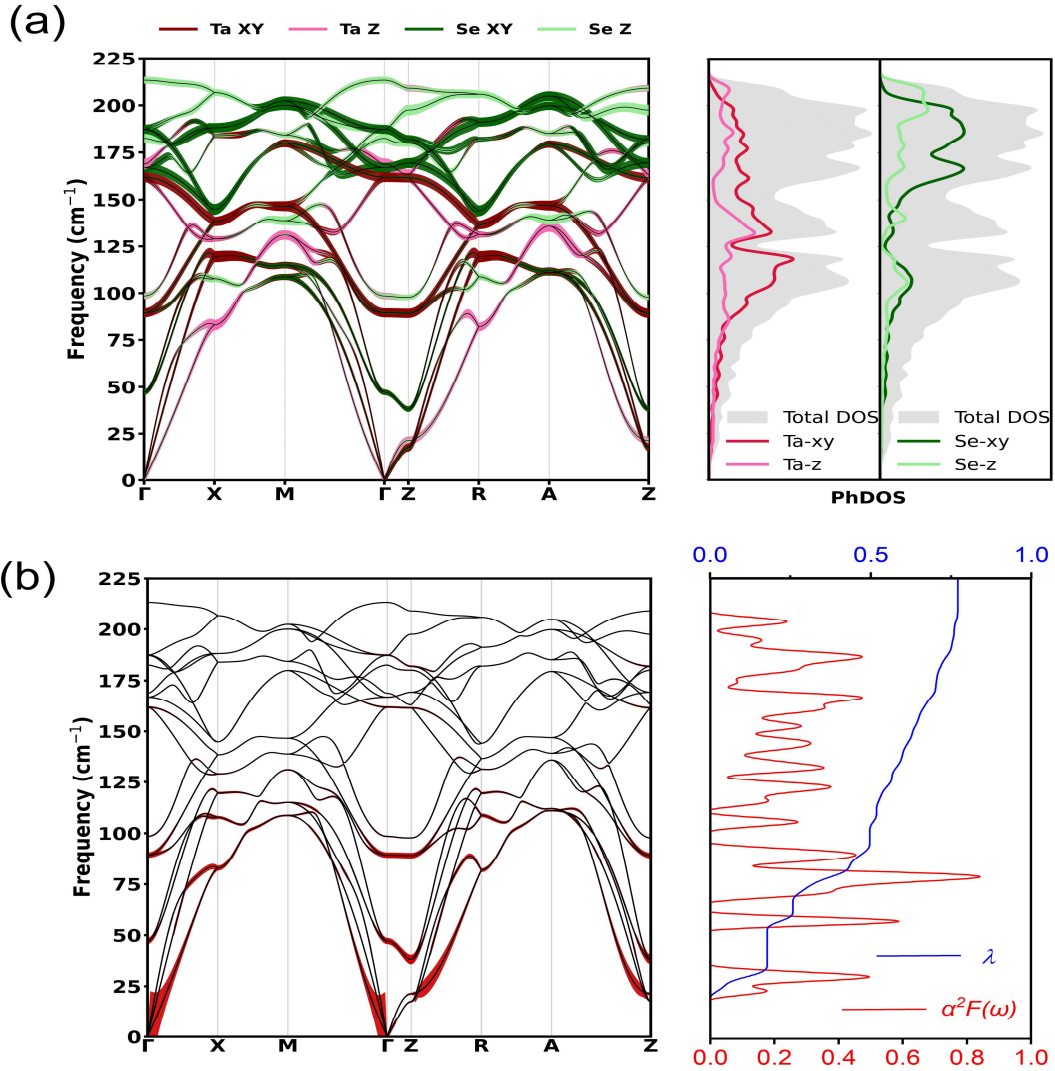

**Figure 20.** (a) Phonon dispersions weighted by different atomic vibrational modes of $Ta_2Se$. The right panel displays the total phonon density of states (gray area) along with the mode-resolved contributions (colored lines). (b) Phonon dispersions of $Ta_2Se$ are shown weighted by the electron-phonon coupling (EPC) strength, along with the Eliashberg spectral function, $\alpha^2F(\omega)$ and the integrated strength of EPC, $\lambda(\omega)$.

The integrated strength of the EPC and the calculated Eliashberg spectral are plotted in **Figure 20**(b). The integrated EPC indicates that phonons below 125 cm$^{-1}$ contribute about 80% of the total coupling for $Ta_2Se$. Although $\alpha^2F(\omega)$ remains finite at higher frequencies (above 125 cm$^{-1}$), its impact on $\lambda$ is smaller, consistent with the weighting in the EPC formula **Equation** (28). The Eliashberg spectral function integrates to average $\lambda$ value of 0.76 which indicates the moderate interaction between electrons and phonons in $Ta_2Se$. Thus, $Ta_2Se$ is weak conventional superconductor.

**Table 10.** Calculated electronic DOS at the Fermi level N($E_F$) (in states/eV. unit cell), strength of EPC ($\lambda$), $\omega_{log}$, and superconducting transition temperature ($T_c$ in K) for Ta$_2$Se.

| Compound | N($E_F$) | $\lambda$ | $\omega_{log}$ | $T_c$ |
|---|---|---|---|---|
| Ta$_2$Se | 1.93 | 0.757 | 117.37 | 3.88 |

The calculated superconducting transition temperature $T_c$, along with the superconducting parameters $\lambda$ and $\omega_{log}$ are listed in **Table 10**. Here, $\mu^*$ represents the electron-electron interaction strength and is typically between 0.10 and 0.16 [116, 117]. In this work, its value is chosen to be 0.13. The computed $T_c$ = 3.88 K matches closely with the experimental value of 3.8 K [12].

The pressure dependence of superconductivity in tetragonal Ta$_2$Se is reflected in the evolution of the density of states at the Fermi level, N($E_F$). With increasing pressure, N($E_F$) decreases monotonically from 3.494 states eV$^{-1}$ at ambient pressure to 2.389 states eV$^{-1}$ at 10 GPa [**Table 9**], indicating a reduction of electronic states near the Fermi energy under compression. Since the electron-phonon coupling strength is proportional to N($E_F$), this decrease weakens the pairing interaction, leading to a suppression of the superconducting transition temperature $T_c$ [115, 116]. The results suggest that pressure-induced band broadening and lattice stiffening dominate the superconducting response of tetragonal Ta$_2$Se up to 10 GPa.

## 4    Conclusions

In this work, tetragonal Ta$_2$Se (P4/*nmm*) is established as a mechanically, thermodynamically, and dynamically stable layered, metal-rich chalcogenide across 0-10 GPa. Hydrostatic pressure drives a monotonic lattice contraction (≈ 9.9% volume reduction by 10 GPa) with stronger compressibility along the c-axis, indicating enhanced interlayer coupling under compression. All independent elastic constants satisfy the Born-Huang stability criteria and increase with pressure, yielding higher *B*, *G* and *Y* and a moderate rise in Vickers hardness. Pugh's ratio and Poisson's ratio consistently indicate ductile, predominantly metallic bonding, while anisotropy measures suggest weak-to-moderate elastic anisotropy that does not signal any instability. Thermophysical descriptors further support a pressure-strengthened lattice. With compression, the mass density increases, the Debye temperature rises, and the melting temperature increases substantially, consistent with stronger bonding and higher vibrational stiffness. The minimum thermal conductivity increases, while the Grüneisen parameter remains within a narrow range, indicating the absence of anomalous anharmonic instabilities under pressure. Bonding analyses (Mulliken populations, bond overlap, and electron-density-difference maps) support a mixed metallic-covalent picture, dominated by a robust Ta-Ta metallic backbone with pressure-enhanced Ta-Se hybridization. Electronic-structure results confirm that Ta$_2$Se remains metallic throughout the pressure range; pressure broadens bands, reduces N($E_F$), reshapes the Fermi surface and suggests a possible Lifshitz-type reconstruction without symmetry breaking. The optical response is consistent with a metal (Drude-like low-energy behavior, high reflectivity), with pressure-tunable spectral features. Phonon dispersions show no imaginary frequencies (0 and 10 GPa), confirming dynamical stability, and EPC calculations classify Ta$_2$Se as a weak-coupling, phonon-mediated superconductor with $T_c$ ≈ 3.9 K, in close agreement with experiment [12].

The marked evolution of $c/c_o$, $N(E_F)$, and vibrational/thermophysical parameters with pressure indicates practical tunability via epitaxial strain or controlled compression in thin films. This provides a clear route to optimize transport, superconducting performance windows, and stability in device environments [6, 118]. The pressure-enhanced stiffness and ductility indicators ($v$, G/B), together with persistent metallic electronic structure, support the use of $Ta_2Se$ as a robust conductive contact material in layered stacks [119, 120]. Its metallic optical response (Drude-like behavior) further motivates exploration as a reflective/EM-screening layer, especially where mechanical robustness and thermal endurance (higher $T_m$) matter [121-124]. Low-temperature superconducting electronics: $Ta_2Se$ can be explored as a ~ 4 K type-II superconducting electrode for Josephson junctions, SQUID loops, and related cryogenic circuit elements, especially where layered geometry/strain-tuning is useful [125, 126]. SOC-enabled proximity platforms: Ta-$d$ dominated bands with strong SOC make $Ta_2Se$ promising as a proximity superconductor in heterostructures aimed at SOC-assisted superconducting and topological device concepts [127, 128]. Future work on tetragonal $Ta_2Se$ should focus on SOC-resolved pressure/strain tuning to test for Lifshitz-type reconstructions, possible band inversion and symmetry-indicated nontrivial topology, complemented by slab calculations to identify any topological surface states.

**Acknowledgements**

S.H.N. acknowledges the research grant (1151/5/52/RU/Science-07/19-20) from the Faculty of Science, University of Rajshahi, Bangladesh, which partly supported this work. This work is dedicated to the cherished memory of the martyrs of the July-August 2024 revolution in Bangladesh, whose sacrifices will forever inspire us.

**Conflict of interest statement**

The authors declare no competing interests.

**Credit author statement**

**Tauhidur Rahman**: Methodology, Software, Formal analysis, Writing- Original draft. **Jubair Hossan Abir**: Methodology, Software, Writing- Reviewing and Editing. **Sourav Kumar Sutradhar**: Methodology, Software, Writing- Reviewing and Editing. **Sraboni Saha Moly**: Methodology, Software, Writing- Reviewing and Editing. **Mst Maskura Khatun**: Methodology, Software, Writing- Reviewing and Editing. **Md. Asif Afzal**: Methodology, Validation, Writing- Reviewing and Editing. **S. H. Naqib**: Conceptualization, Supervision, Formal Analysis, Writing- Reviewing and Editing.

**Data availability statement**

The data sets generated and/or analyzed in this study are available from the corresponding author upon reasonable request.

**List of References**

[1] T. Hughbanks, "Exploring the metal-rich chemistry of the early transition elements," *Journal of Alloys and Compounds*, vol. 229, no. 1, pp. 40–53, Oct. 1995, doi: 10.1016/0925-8388(95)01688-0.

[2] L. Chen, S.-Q. Xia, and J. D. Corbett, "Metal-Rich Chalcogenides. Synthesis, Structure, and Bonding of the Layered $Lu_{11}Te_4$. Comparison with the Similar $Sc_8Te_3$ and $Ti_{11}Se_4$," *Inorganic Chemistry*, vol. 44, no. 9, pp. 3057–3062, May 2005, doi: 10.1021/ic0401142.

[3] J. F. Khoury *et al.*, "$Ir_6In_{32}S_{21}$, a polar, metal-rich semiconducting subchalcogenide," *Chemical Science*, vol. 11, no. 3, pp. 870–878, 2020.

[4] H. Wang, H. Yuan, S. Sae Hong, Y. Li, and Y. Cui, "Physical and chemical tuning of two-dimensional transition metal dichalcogenides," *Chemical Society Reviews*, vol. 44, no. 9, pp. 2664–2680, 2015, doi: 10.1039/C4CS00287C.

[5] S. Manzeli, D. Ovchinnikov, D. Pasquier, O. V. Yazyev, and A. Kis, "2D transition metal dichalcogenides," *Nature Reviews Materials*, vol. 2, no. 8, p. 17033, Jun. 2017, doi: 10.1038/natrevmats.2017.33.

[6] Y. Qi *et al.*, "Recent Progress in Strain Engineering on Van der Waals 2D Materials: Tunable Electrical, Electrochemical, Magnetic, and Optical Properties," *Advanced Materials*, vol. 35, no. 12, p. 2205714, Mar. 2023, doi: 10.1002/adma.202205714.

[7] G. J. Snyder and E. S. Toberer, "Complex thermoelectric materials," *Nature Materials*, vol. 7, no. 2, pp. 105–114, Feb. 2008, doi: 10.1038/nmat2090.

[8] S. Puthran, G. S. Hegde, and A. N. Prabhu, "Review of Chalcogenide-Based Materials for Low-, Mid-, and High-Temperature Thermoelectric Applications," *Journal of Electronic Materials*, vol. 53, no. 10, pp. 5739–5768, Oct. 2024, doi: 10.1007/s11664-024-11310-7.

[9] P. Muthu *et al.*, "Review of Transition Metal Chalcogenides and Halides as Electrode Materials for Thermal Batteries and Secondary Energy Storage Systems," *ACS Omega*, vol. 9, no. 7, pp. 7357–7374, Feb. 2024, doi: 10.1021/acsomega.3c08809.

[10] P. Canepa *et al.*, "Odyssey of Multivalent Cathode Materials: Open Questions and Future Challenges," *Chemical Reviews*, vol. 117, no. 5, pp. 4287–4341, Mar. 2017, doi: 10.1021/acs.chemrev.6b00614.

[11] B. Harbrecht, "$Ta_2Se$: Ein tantalreiches Selenid mit einer neuen Schichtstruktur," *Angewandte Chemie*, vol. 101, no. 12, pp. 1696–1698, Dec. 1989, doi: 10.1002/ange.19891011211.

[12] X. Gui, K. Górnicka, Q. Chen, H. Zhou, T. Klimczuk, and W. Xie, "Superconductivity in Metal-Rich Chalcogenide $Ta_2Se$," *Inorganic Chemistry*, vol. 59, no. 9, pp. 5798–5802, May 2020, doi: 10.1021/acs.inorgchem.9b03656.

[13] J. Lee *et al.*, "Identification of phonon vibrational modes for the layered $Ta_2Se$ metal-rich chalcogenide," *npj 2D Materials and Applications*, Nov. 2025, doi: 10.1038/s41699-025-00632-7.

[14] J. V. Yakhmi, *Superconducting Materials and Their Applications: An Interdisciplinary Approach*, 1st ed. in IOP Ebooks Series. Bristol: Institute of Physics Publishing, 2021.

[15] G. Marini, A. Sanna, C. Pellegrini, C. Bersier, E. Tosatti, and G. Profeta, "Superconducting Chevrel phase $PbMo_6S_8$ from first principles," *Physical Review B*, vol. 103, no. 14, p. 144507, Apr. 2021, doi: 10.1103/PhysRevB.103.144507.

[16] L. G. Pimenta Martins, R. Comin, M. J. S. Matos, M. S. C. Mazzoni, B. R. A. Neves, and M. Yankowitz, "High-pressure studies of atomically thin van der Waals materials," *Applied Physics Reviews*, vol. 10, no. 1, p. 011313, Mar. 2023, doi: 10.1063/5.0123283.

[17] D. C. Freitas *et al.*, "Strong enhancement of superconductivity at high pressures within the charge-density-wave states of $2H-TaS_2$ and $2H-TaSe_2$," *Physical Review B*, vol. 93, no. 18, p. 184512, May 2016, doi: 10.1103/PhysRevB.93.184512.

[18] S. Xu *et al.*, "Effects of disorder and hydrostatic pressure on charge density wave and superconductivity in 2H-TaS$_2$," *Physical Review B*, vol. 103, no. 22, p. 224509, Jun. 2021, doi: 10.1103/PhysRevB.103.224509.

[19] Y. Tymoshenko *et al.*, "Charge-density-wave quantum critical point under pressure in 2H-TaSe$_2$," *Communications Physics*, vol. 8, no. 1, p. 352, Aug. 2025, doi: 10.1038/s42005-025-02254-3.

[20] M. Güler, Ş. Uğur, G. Uğur, and E. Güler, "First principles study of the electronic, optical, elastic and thermoelectric properties of Nb$_2$WNi alloy," *Molecular Physics*, vol. 119, no. 12, p. e1928314, Jun. 2021, doi: 10.1080/00268976.2021.1928314.

[21] R. G. Parr, "Density Functional Theory," *Annual Review of Physical Chemistry*, vol. 34, no. 1, pp. 631–656, Oct. 1983, doi: 10.1146/annurev.pc.34.100183.003215.

[22] S. Clark *et al.*, "First principles methods using CASTEP," *Zeitschrift für Kristallographie*, vol. 220, May 2005, doi: 10.1524/zkri.220.5.567.65075.

[23] E. Güler, M. Güler, Ş. Uğur, and G. Uğur, "First principles study of electronic, elastic, optical and magnetic properties of Rh$_2$MnX (X = Ti, Hf, Sc, Zr, Zn) Heusler alloys," *International Journal of Quantum Chemistry*, vol. 121, no. 10, p. e26606, 2021, doi: 10.1002/qua.26606.

[24] D. Vanderbilt, "Soft self-consistent pseudopotentials in a generalized eigenvalue formalism," *Physical Review B*, vol. 41, no. 11, pp. 7892–7895, Apr. 1990, doi: 10.1103/PhysRevB.41.7892.

[25] G. Kresse and J. Furthmüller, "Efficiency of ab-initio total energy calculations for metals and semiconductors using a plane-wave basis set," *Computational Materials Science*, vol. 6, no. 1, pp. 15–50, Jul. 1996, doi: 10.1016/0927-0256(96)00008-0.

[26] G. Kresse and J. Furthmüller, "Efficient iterative schemes for ab initio total-energy calculations using a plane-wave basis set," *Physical Review B*, vol. 54, no. 16, pp. 11169–11186, Oct. 1996, doi: 10.1103/PhysRevB.54.11169.

[27] P. E. Blöchl, "Projector augmented-wave method," *Physical Review B*, vol. 50, no. 24, pp. 17953–17979, Dec. 1994, doi: 10.1103/PhysRevB.50.17953.

[28] G. Kresse and D. Joubert, "From ultrasoft pseudopotentials to the projector augmented-wave method," *Physical Review B*, vol. 59, no. 3, pp. 1758–1775, Jan. 1999, doi: 10.1103/PhysRevB.59.1758.

[29] J. P. Perdew *et al.*, "Restoring the Density-Gradient Expansion for Exchange in Solids and Surfaces," *Physical Review Letters*, vol. 100, no. 13, p. 136406, Apr. 2008, doi: 10.1103/PhysRevLett.100.136406.

[30] T. H. Fischer and J. Almlof, "General methods for geometry and wave function optimization," *The Journal of Physical Chemistry*, vol. 96, no. 24, pp. 9768–9774, Nov. 1992, doi: 10.1021/j100203a036.

[31] H. J. Monkhorst and J. D. Pack, "Special points for Brillouin-zone integrations," *Physical Review B*, vol. 13, no. 12, pp. 5188–5192, Jun. 1976, doi: 10.1103/PhysRevB.13.5188.

[32] G. P. Francis and M. C. Payne, "Finite basis set corrections to total energy pseudopotential calculations," *Journal of Physics: Condensed Matter*, vol. 2, no. 19, pp. 4395–4404, May 1990, doi: 10.1088/0953-8984/2/19/007.

[33] O. H. Nielsen and R. M. Martin, "First-Principles Calculation of Stress," *Physical Review Letters*, vol. 50, no. 9, pp. 697–700, Feb. 1983, doi: 10.1103/PhysRevLett.50.697.

[34] J. P. Watt, "Hashin-Shtrikman bounds on the effective elastic moduli of polycrystals with orthorhombic symmetry," *Journal of Applied Physics*, vol. 50, no. 10, pp. 6290–6295, Oct. 1979, doi: 10.1063/1.325768.

[35] J. P. Watt and L. Peselnick, "Clarification of the Hashin-Shtrikman bounds on the effective elastic moduli of polycrystals with hexagonal, trigonal, and tetragonal symmetries,"

*Journal of Applied Physics*, vol. 51, no. 3, pp. 1525–1531, Mar. 1980, doi: 10.1063/1.327804.

[36] F. Subhan *et al.*, "Elastic and optoelectronic properties of CaTa$_2$O$_6$ compounds: Cubic and orthorhombic phases," *Journal of Alloys and Compounds*, vol. 785, pp. 232–239, May 2019, doi: 10.1016/j.jallcom.2019.01.140.

[37] M. A. Ghebouli *et al.*, "Electronic band structure, elastic, optical and thermodynamic characteristic of cubic YF$_3$: An *ab-initio* study," *Optik*, vol. 239, p. 166680, Aug. 2021, doi: 10.1016/j.ijleo.2021.166680.

[38] D. Sanchez-Portal, E. Artacho, and J. M. Soler, "Projection of plane-wave calculations into atomic orbitals," *Solid State Communications*, vol. 95, no. 10, pp. 685–690, Sept. 1995, doi: 10.1016/0038-1098(95)00341-X.

[39] M. D. Segall, R. Shah, C. J. Pickard, and M. C. Payne, "Population analysis of plane-wave electronic structure calculations of bulk materials," *Physical Review B*, vol. 54, no. 23, pp. 16317–16320, Dec. 1996, doi: 10.1103/PhysRevB.54.16317.

[40] R. S. Mulliken, "Electronic Population Analysis on LCAO–MO Molecular Wave Functions. II. Overlap Populations, Bond Orders, and Covalent Bond Energies," *The Journal of Chemical Physics*, vol. 23, no. 10, pp. 1841–1846, Oct. 1955, doi: 10.1063/1.1740589.

[41] F. Birch, "Finite strain isotherm and velocities for single-crystal and polycrystalline NaCl at high pressures and 300°K," *Journal of Geophysical Research: Solid Earth*, vol. 83, no. B3, pp. 1257–1268, Mar. 1978, doi: 10.1029/JB083iB03p01257.

[42] P. Giannozzi *et al.*, "QUANTUM ESPRESSO: a modular and open-source software project for quantum simulations of materials," *Journal of Physics: Condensed Matter*, vol. 21, no. 39, p. 395502, Sept. 2009, doi: 10.1088/0953-8984/21/39/395502.

[43] A. Dal Corso, "Pseudopotentials periodic table: From H to Pu," *Computational Materials Science*, vol. 95, pp. 337–350, Dec. 2014, doi: 10.1016/j.commatsci.2014.07.043.

[44] S. H. Naqib, M. T. Hoque, and A. K. M. A. Islam, "Oxygen depletion dependence of pressure coefficient of YBCO(123)," in *Advances in high pressure science and technology: proceedings of the fourth national conference on high pressure science and technology*, 1997, pp. 168–173. Accessed: Dec. 16, 2025. [Online]. Available: https://inis.iaea.org/records/xtr91-a1r27

[45] R. Islam, S. Naqib, and A. K. M. Islam, "Modeling of the Complex Doping Dependence of dT$_c$/dP of YBa$_2$Cu$_3$O$_{6+x}$," *Journal of Superconductivity*, vol. 13, pp. 485–490, Jan. 2000, doi: 10.1023/A:1007775630383.

[46] M. I. Naher and S. H. Naqib, "Structural, elastic, electronic, bonding, and optical properties of topological CaSn$_3$ semimetal," *Journal of Alloys and Compounds*, vol. 829, p. 154509, Jul. 2020, doi: 10.1016/j.jallcom.2020.154509.

[47] C. Sammis, "Thermodynamics of crystals by DC Wallace," *Foundations of Crystallography*, vol. 29, no. 5, pp. 582–583, 1973.

[48] D. G. Pettifor, "Theoretical predictions of structure and related properties of intermetallics," *Materials Science and Technology*, vol. 8, no. 4, pp. 345–349, Apr. 1992, doi: 10.1179/mst.1992.8.4.345.

[49] L. Kleinman, "Deformation Potentials in Silicon. I. Uniaxial Strain," *Physical Review*, vol. 128, no. 6, pp. 2614–2621, Dec. 1962, doi: 10.1103/PhysRev.128.2614.

[50] M. Jamal, S. Jalali Asadabadi, I. Ahmad, and H. Aliabad, "Elastic constants of cubic crystals," *Computational Materials Science*, vol. 95, p. 592, Aug. 2014, doi: 10.1016/j.commatsci.2014.08.027.

[51] A. Gueddouh, B. Bentria, and I. K. Lefkaier, "First-principle investigations of structure, elastic and bond hardness of Fe$_x$B (x=1, 2, 3) under pressure," *Journal of Magnetism and Magnetic Materials*, vol. 406, pp. 192–199, May 2016, doi: 10.1016/j.jmmm.2016.01.013.

[52] M. Mattesini, R. Ahuja, and B. Johansson, "Cubic Hf$_3$N$_4$ and Zr$_3$N$_4$: A class of hard materials," *Physical Review B*, vol. 68, no. 18, p. 184108, Nov. 2003, doi: 10.1103/PhysRevB.68.184108.

[53] J. A. Majewski and P. Vogl, "Simple model for structural properties and crystal stability of sp-bonded solids," *Physical Review B*, vol. 35, no. 18, pp. 9666–9682, Jun. 1987, doi: 10.1103/PhysRevB.35.9666.

[54] P. Ravindran, L. Fast, P. A. Korzhavyi, B. Johansson, J. Wills, and O. Eriksson, "Density functional theory for calculation of elastic properties of orthorhombic crystals: Application to TiSi$_2$," *Journal of Applied Physics*, vol. 84, no. 9, pp. 4891–4904, Nov. 1998, doi: 10.1063/1.368733.

[55] C. Kittel, *Introduction to solid state physics*, 8. ed., [Repr.]. Hoboken, NJ: Wiley, 20.

[56] S. F. Pugh, "XCII. Relations between the elastic moduli and the plastic properties of polycrystalline pure metals," *The London, Edinburgh, and Dublin Philosophical Magazine and Journal of Science*, vol. 45, no. 367, pp. 823–843, Aug. 1954, doi: 10.1080/14786440808520496.

[57] V. V. Bannikov, I. R. Shein, and A. L. Ivanovskii, "Elastic properties of antiperovskite-type Ni-rich nitrides MNNi$_3$(M=Zn, Cd, Mg, Al, Ga, In, Sn, Sb, Pd, Cu, Ag and Pt) as predicted from first-principles calculations," *Physica B: Condensed Matter*, vol. 405, no. 22, pp. 4615–4619, Nov. 2010, doi: 10.1016/j.physb.2010.08.046.

[58] W. Voigt, *Lehrbuch der kristallphysik (mit ausschluss der kristalloptik)*. Leipzig, Berlin, B.G. Teubner, 1910. Accessed: Dec. 07, 2025. [Online]. Available: http://archive.org/details/bub_gb_SvPPAAAAMAAJ

[59] P. H. Mott, J. R. Dorgan, and C. M. Roland, "The bulk modulus and Poisson's ratio of 'incompressible' materials," *Journal of Sound and Vibration*, vol. 312, no. 4, pp. 572–575, May 2008, doi: 10.1016/j.jsv.2008.01.026.

[60] A. Šimůnek, "How to estimate hardness of crystals on a pocket calculator," *Physical Review B*, vol. 75, no. 17, p. 172108, May 2007, doi: 10.1103/PhysRevB.75.172108.

[61] Z. Sun, D. Music, R. Ahuja, and J. M. Schneider, "Theoretical investigation of the bonding and elastic properties of nanolayered ternary nitrides," *Physical Review B*, vol. 71, no. 19, p. 193402, May 2005, doi: 10.1103/PhysRevB.71.193402.

[62] M. J. Phasha, P. E. Ngoepe, H. R. Chauke, D. G. Pettifor, and D. Nguyen-Mann, "Link between structural and mechanical stability of fcc- and bcc-based ordered Mg–Li alloys," *Intermetallics*, vol. 18, no. 11, pp. 2083–2089, Nov. 2010, doi: 10.1016/j.intermet.2010.06.015.

[63] R. C. Lincoln, K. M. Koliwad, and P. B. Ghate, "Morse-Potential Evaluation of Second- and Third-Order Elastic Constants of Some Cubic Metals," *Physical Review*, vol. 157, no. 3, pp. 463–466, May 1967, doi: 10.1103/PhysRev.157.463.

[64] L. Vitos, P. A. Korzhavyi, and B. Johansson, "Stainless steel optimization from quantum mechanical calculations," *Nature Materials*, vol. 2, no. 1, pp. 25–28, Jan. 2003, doi: 10.1038/nmat790.

[65] M. A. Ali, M. M. Hossain, A. K. M. A. Islam, and S. H. Naqib, "Ternary boride Hf$_3$PB$_4$: Insights into the physical properties of the hardest possible boride MAX phase," *Journal of Alloys and Compounds*, vol. 857, p. 158264, Mar. 2021, doi: 10.1016/j.jallcom.2020.158264.

[66] M. I. Naher and S. H. Naqib, "An ab-initio study on structural, elastic, electronic, bonding, thermal, and optical properties of topological Weyl semimetal TaX (X = P, As)," *Scientific Reports*, vol. 11, no. 1, p. 5592, Mar. 2021, doi: 10.1038/s41598-021-85074-z.

[67] C. M. Kube, "Elastic anisotropy of crystals," *AIP Advances*, vol. 6, no. 9, Sept. 2016, doi: 10.1063/1.4962996.

[68] M. Islam, M. Sarker, Y. Inagaki, and S. Islam, "Study of a new layered ternary chalcogenide CuZnTe$_2$ and its potassium intercalation effect," *Materials Research Express*, vol. 7, p. 105904, Oct. 2020, doi: 10.1088/2053-1591/abbd06.

[69] Y. Al-Khatatbeh and K. K. M. Lee, "From superhard to hard: A review of transition metal dioxides TiO$_2$, ZrO$_2$, and HfO$_2$ hardness," *Journal of Superhard Materials*, vol. 36, no. 4, pp. 231–245, Jul. 2014, doi: 10.3103/S1063457614040029.

[70] S. I. Ranganathan and M. Ostoja-Starzewski, "Universal Elastic Anisotropy Index," *Physical Review Letters*, vol. 101, no. 5, p. 055504, Aug. 2008, doi: 10.1103/PhysRevLett.101.055504.

[71] D. H. Chung and W. R. Buessem, "The Elastic Anisotropy of Crystals," *Journal of Applied Physics*, vol. 38, no. 5, pp. 2010–2012, Apr. 1967, doi: 10.1063/1.1709819.

[72] C. M. Kube and M. De Jong, "Elastic constants of polycrystals with generally anisotropic crystals," *Journal of Applied Physics*, vol. 120, no. 16, p. 165105, Oct. 2016, doi: 10.1063/1.4965867.

[73] V. Milman and M. C. Warren, "Elasticity of hexagonal BeO," *Journal of Physics: Condensed Matter*, vol. 13, no. 2, pp. 241–251, Jan. 2001, doi: 10.1088/0953-8984/13/2/302.

[74] R. Gaillac, P. Pullumbi, and F.-X. Coudert, "ELATE: an open-source online application for analysis and visualization of elastic tensors," *Journal of Physics: Condensed Matter*, vol. 28, no. 27, p. 275201, May 2016, doi: 10.1088/0953-8984/28/27/275201.

[75] R. S. Krishnan, R. Srinivasan, and S. Devanarayanan, *Thermal Expansion of Crystals: International Series in the Science of the Solid State*. Pergamon, 2013.

[76] O. L. Anderson, "A simplified method for calculating the debye temperature from elastic constants," *Journal of Physics and Chemistry of Solids*, vol. 24, no. 7, pp. 909–917, Jul. 1963, doi: 10.1016/0022-3697(63)90067-2.

[77] F. Parvin and S. H. Naqib, "Pressure dependence of structural, elastic, electronic, thermodynamic, and optical properties of van der Waals-type NaSn$_2$P$_2$ pnictide superconductor: Insights from DFT study," *Results in Physics*, vol. 21, p. 103848, Feb. 2021, doi: 10.1016/j.rinp.2021.103848.

[78] W. Kim, "Strategies for engineering phonon transport in thermoelectrics," *Journal of Materials Chemistry C*, vol. 3, no. 40, pp. 10336–10348, 2015, doi: 10.1039/C5TC01670C.

[79] M. I. Naher and S. H. Naqib, "A comprehensive study of the thermophysical and optoelectronic properties of Nb$_2$P$_5$ via ab-initio technique," *Results in Physics*, vol. 28, p. 104623, Sept. 2021, doi: 10.1016/j.rinp.2021.104623.

[80] M. I. Naher and S. H. Naqib, "Possible applications of Mo$_2$C in the orthorhombic and hexagonal phases explored via *ab-initio* investigations of elastic, bonding, optoelectronic and thermophysical properties," *Results in Physics*, vol. 37, p. 105505, Jun. 2022, doi: 10.1016/j.rinp.2022.105505.

[81] M. Kholil, M. Rahaman, and Md. A. Rahman, "First principles study of the structural, elastic, electronic, optical and thermodynamic properties of SrRh$_2$ laves phase intermetallic compound," *Computational Condensed Matter*, vol. 13, pp. 65–71, Sept. 2017, doi: 10.1016/j.cocom.2017.09.008.

[82] D. Clarke, "Materials Selection Guidelines for Low Thermal Conductivity TBCs," *Surface & Coatings Technology - SURF COAT TECH*, vol. 163, pp. 67–74, Jan. 2003, doi: 10.1016/S0257-8972(02)00593-5.

[83] B. D. Sanditov, Sh. B. Tsydypov, and D. S. Sanditov, "Relation between the Grüneisen constant and Poisson's ratio of vitreous systems," *Acoustical Physics*, vol. 53, no. 5, pp. 594–597, Sept. 2007, doi: 10.1134/S1063771007050090.

[84] R. E. Taylor, *Thermal Expansion of Solids*. ASM International, 1998.

[85] M. I. Naher, M. A. Ali, M. M. Hossain, M. M. Uddin, and S. H. Naqib, "A comprehensive *ab-initio* insights into the pressure dependent mechanical, phonon, bonding, electronic, optical, and thermal properties of $CsV_3Sb_5$ Kagome compound," *Results in Physics*, vol. 51, p. 106742, Aug. 2023, doi: 10.1016/j.rinp.2023.106742.

[86] J. Li, N. V. Medhekar, and V. B. Shenoy, "Bonding Charge Density and Ultimate Strength of Monolayer Transition Metal Dichalcogenides," *The Journal of Physical Chemistry C*, vol. 117, no. 30, pp. 15842–15848, Aug. 2013, doi: 10.1021/jp403986v.

[87] S. Li, R. Ahuja, M. W. Barsoum, P. Jena, and B. Johansson, "Optical properties of $Ti_3SiC_2$ and $Ti_4AlN_3$," *Applied Physics Letters*, vol. 92, no. 22, p. 221907, Jun. 2008, doi: 10.1063/1.2938862.

[88] H. Wang, Y. Chen, Y. Kaneta, and S. Iwata, "First-principles study on effective doping to improve the optical properties in spinel nitrides," *Journal of Alloys and Compounds*, vol. 491, no. 1–2, pp. 550–559, Feb. 2010, doi: 10.1016/j.jallcom.2009.10.267.

[89] S. Naderizadeh, S. M. Elahi, M. R. Abolhassani, F. Kanjouri, N. Rahimi, and J. Jalilian, "Electronic and optical properties of Full-Heusler alloy $Fe_{3-x}Mn_xSi$," *The European Physical Journal B*, vol. 85, no. 5, p. 144, May 2012, doi: 10.1140/epjb/e2012-20919-3.

[90] M. Fox, *Optical properties of solids*, Second edition, Reprinted (with corrections). in Oxford master series in condensed matter physics, no. 3. Oxford New York, NY: Oxford University Press, 2011.

[91] J. A. Wilson and A. D. Yoffe, "The transition metal dichalcogenides discussion and interpretation of the observed optical, electrical and structural properties," *Advances in Physics*, vol. 18, no. 73, pp. 193–335, May 1969, doi: 10.1080/00018736900101307.

[92] M. Qi *et al.*, "Pressure-driven Lifshitz transition in type-II Dirac semimetal $NiTe_2$," *Physical Review B*, vol. 101, no. 11, p. 115124, Mar. 2020, doi: 10.1103/PhysRevB.101.115124.

[93] H. Yang *et al.*, "Pressure-induced nontrivial $Z_2$ band topology and superconductivity in the transition metal chalcogenide $Ta_2Ni_3Te_5$," *Physical Review B*, vol. 107, no. 2, p. L020503, Jan. 2023, doi: 10.1103/PhysRevB.107.L020503.

[94] K. Boudiaf, A. Bouhemadou, Y. Al-Douri, R. Khenata, S. Bin-Omran, and N. Guechi, "Electronic and thermoelectric properties of the layered BaFAgCh(Ch = S, Se and Te): First-principles study," *Journal of Alloys and Compounds*, vol. 759, pp. 32–43, Aug. 2018, doi: 10.1016/j.jallcom.2018.05.142.

[95] A. Bekhti Siad *et al.*, "Electronic, optical and thermoelectric investigations of Zintl phase $AE_3AlAs_3$(AE= Sr, Ba): First-principles calculations," *Chinese Journal of Physics*, vol. 56, Mar. 2018, doi: 10.1016/j.cjph.2018.03.022.

[96] A. Belhachemi, H. Abid, Y. Al-Douri, M. Sehil, A. Bouhemadou, and M. Ameri, "First-principles calculations to investigate the structural, electronic and optical properties of $Zn_{1-x}Mg_xTe$ ternary alloys," *Chinese Journal of Physics*, vol. 55, no. 3, pp. 1018–1031, Jun. 2017, doi: 10.1016/j.cjph.2017.02.018.

[97] K. S. Nanjundaswamy and T. Hughbanks, "Subtleties of structure and bonding in Ta−S−Se and Ta−Nb−S solid solutions," *Journal of Solid State Chemistry*, vol. 98, no. 2, pp. 278–290, Jun. 1992, doi: 10.1016/S0022-4596(05)80236-1.

[98] C. L. Kane and E. J. Mele, "Quantum Spin Hall Effect in Graphene," *Physical Review Letters*, vol. 95, no. 22, p. 226801, Nov. 2005, doi: 10.1103/PhysRevLett.95.226801.

[99] J.-H. Xu, T. Oguchi, and A. J. Freeman, "Solid-solution strengthening: Substitution of V in $Ni_3Al$ and structural stability of $Ni_3(Al,V)$," *Physical Review B*, vol. 36, no. 8, pp. 4186–4189, Sept. 1987, doi: 10.1103/PhysRevB.36.4186.

[100] T. Hong, T. J. Watson-Yang, A. J. Freeman, T. Oguchi, and J. Xu, "Crystal structure, phase stability, and electronic structure of Ti-Al intermetallics: $TiAl_3$," *Physical Review B*, vol. 41, no. 18, pp. 12462–12467, Jun. 1990, doi: 10.1103/PhysRevB.41.12462.

[101] A. Pasturel, C. Colinet, and P. Hicter, "Strong chemical interactions in disordered alloys," *Physica B+C*, vol. 132, no. 2, pp. 177–180, Jul. 1985, doi: 10.1016/0378-4363(85)90062-2.

[102] I. Galanakis and P. Mavropoulos, "Spin-polarization and electronic properties of half-metallic Heusler alloys calculated from first principles," *Journal of Physics: Condensed Matter*, vol. 19, no. 31, p. 315213, Jul. 2007, doi: 10.1088/0953-8984/19/31/315213.

[103] W. Lin, J. Xu, and A. J. Freeman, "Electronic structure, cohesive properties, and phase stability of $Ni_3V$, $Co_3V$, and $Fe_3V$," *Physical Review B*, vol. 45, no. 19, pp. 10863–10871, May 1992, doi: 10.1103/PhysRevB.45.10863.

[104] V. L. Moruzzi, P. Oelhafen, A. R. Williams, R. Lapka, H.-J. Güntherodt, and J. Kübler, "Theoretical and experimental electronic structure of Zr-based transition-metal glasses containing Fe, Co, Ni, Cu, Rh, and Pd," *Physical Review B*, vol. 27, no. 4, pp. 2049–2054, Feb. 1983, doi: 10.1103/PhysRevB.27.2049.

[105] D. H. Douglass, *Superconductivity in d- and f- band metals; Proceedings of the Conference, University of Rochester, Rochester, N.Y., October 29, 30, 1971*. 1972. Accessed: Dec. 16, 2025. [Online]. Availabl e: https://ntrs.nasa.gov/citations/19720038885

[106] G. Jency and S. Kumari, "*ab Initio* Study on the Electronic Band Structure, Density of States, Structural Phase Transition and Superconductivity of Zirconium," *Universal Journal of Chemistry*, vol. 1, pp. 64–70, Aug. 2013, doi: 10.13189/ujc.2013.010207.

[107] M. Hadi, S. Naqib, S.-R. G. Christopoulos, A. Chroneos, and A. K. M. Islam, "Mechanical behavior, bonding nature and defect processes of $Mo_2ScAlC_2$: A new ordered MAX phase," *Journal of Alloys and Compounds*, vol. 724, Jul. 2017, doi: 10.1016/j.jallcom.2017.07.110.

[108] A. Chowdhury, M. A. Ali, M. M. Hossain, M. M. Uddin, S. H. Naqib, and A. K. M. A. Islam, "Predicted MAX Phase $Sc_2InC$: Dynamical Stability, Vibrational and Optical Properties," *physica status solidi (b)*, vol. 255, no. 3, p. 1700235, 2018, doi: 10.1002/pssb.201700235.

[109] P. Cudazzo *et al.*, "*ab-Initio* Description of High-Temperature Superconductivity in Dense Molecular Hydrogen," *Physical Review Letters*, vol. 100, no. 25, p. 257001, Jun. 2008, doi: 10.1103/PhysRevLett.100.257001.

[110] E. G. Michel, "Fermi surface analysis using surface methods," *Journal of Physics: Condensed Matter*, vol. 19, no. 35, p. 350301, Sept. 2007, doi: 10.1088/0953-8984/19/35/350301.

[111] G. Kresse, J. Furthmüller, and J. Hafner, "Ab initio force constant approach to phonon dispersion relations of diamond and graphite," *Europhysics Letters*, vol. 32, no. 9, p. 729, 1995.

[112] E. N. Koukaras, G. Kalosakas, C. Galiotis, and K. Papagelis, "Phonon properties of graphene derived from molecular dynamics simulations," *Scientific reports*, vol. 5, no. 1, p. 12923, 2015.

[113] K. Parlinski, Z. Q. Li, and Y. Kawazoe, "First-Principles Determination of the Soft Mode in Cubic $ZrO_2$," *Physical Review Letters*, vol. 78, no. 21, pp. 4063–4066, May 1997, doi: 10.1103/PhysRevLett.78.4063.

[114] V. Heine, "*Group Theory: Application to the Physics of Condensed Matter*," *Physics Today*, vol. 61, no. 11, pp. 57–58, Nov. 2008, doi: 10.1063/1.3027994.

[115] W. L. McMillan, "Transition Temperature of Strong-Coupled Superconductors," *Physical Review*, vol. 167, no. 2, pp. 331–344, Mar. 1968, doi: 10.1103/PhysRev.167.331.

[116] P. B. Allen and R. C. Dynes, "Transition temperature of strong-coupled superconductors reanalyzed," *Physical Review B*, vol. 12, no. 3, pp. 905–922, Aug. 1975, doi: 10.1103/PhysRevB.12.905.

[117] P. B. Allen and R. C. Dynes, "Superconductivity at very strong coupling," *Journal of Physics C: Solid State Physics*, vol. 8, no. 9, pp. L158–L163, May 1975, doi: 10.1088/0022-3719/8/9/020.

[118] Y. Chernolevska *et al.*, "Strain-controlled superconductivity in epitaxially grown thin films of 1T-TaS$_2$," *Scientific Reports*, vol. 15, no. 1, p. 36052, Oct. 2025, doi: 10.1038/s41598-025-19901-y.

[119] O. N. Senkov and D. B. Miracle, "Generalization of intrinsic ductile-to-brittle criteria by Pugh and Pettifor for materials with a cubic crystal structure," *Scientific reports*, vol. 11, no. 1, p. 4531, Feb. 2021, doi: 10.1038/s41598-021-83953-z.

[120] S. Kojima, "Poisson's Ratio of Glasses, Ceramics, and Crystals," *Materials*, vol. 17, no. 2, p. 300, Jan. 2024, doi: 10.3390/ma17020300.

[121] G. Grüner and M. Dressel, Eds., *Electrodynamics of solids: optical properties of electrons in matter*. Cambridge New York: Cambridge University Press, 2002.

[122] H. Zhao *et al.*, "Electromagnetic Interference Shielding Films: Structure Design and Prospects," *Small Methods*, vol. 9, no. 5, p. 2401324, 2025, doi: 10.1002/smtd.202401324.

[123] R. L. Fagaly, "Superconducting quantum interference device instruments and applications," *Review of Scientific Instruments*, vol. 77, no. 10, p. 101101, Oct. 2006, doi: 10.1063/1.2354545.

[124] L. You, "Superconducting nanowire single-photon detectors for quantum information," *Nanophotonics*, vol. 9, no. 9, pp. 2673–2692, 2020.

[125] A. I. Braginski and J. Clarke, Eds., *The SQUID handbook: Vol. 1: Fundamentals and technology of SQUIDs and SQUID systems*. Weinheim: Wiley-VCH, 2004. doi: 10.1002/3527603646.

[126] Ya. S. Greenberg, "Application of superconducting quantum interference devices to nuclear magnetic resonance," *Reviews of Modern Physics*, vol. 70, no. 1, pp. 175–222, Jan. 1998, doi: 10.1103/RevModPhys.70.175.

[127] C. Trang *et al.*, "Conversion of a conventional superconductor into a topological superconductor by topological proximity effect," *Nature communications*, vol. 11, no. 1, p. 159, 2020.

[128] V. Mishra, Y. Li, F.-C. Zhang, and S. Kirchner, "Effects of spin-orbit coupling on proximity-induced superconductivity," *Physical Review B*, vol. 107, no. 18, p. 184505, 2023.